%% file: CD-Paper.tex
\documentclass[sn-aps]{sn-jnl}

\usepackage{xpatch}
\makeatletter
\xpatchcmd{\@@author}{\ifx#1}{\ifx{#1}}{}{}
\makeatother

\usepackage{graphicx}
\usepackage{xcolor}
\usepackage{caption}
\usepackage{subcaption}
\usepackage{multirow}
\usepackage[numbers,sort&compress]{natbib}
\usepackage{threeparttable}
\usepackage{textcomp}
\usepackage{tabularx}
\usepackage{pifont}

\title{The Design and Technology Development of the JUNO Central Detector}

\include{fullname_231009}

\date{Received: date / Accepted: date}

\abstract{The Jiangmen Underground Neutrino Observatory (JUNO) is a large-scale neutrino experiment with multiple physics goals including determining the neutrino mass hierarchy, the accurate measurement of neutrino oscillation parameters, the neutrino detection from the supernova, the Sun, and the Earth, etc. JUNO puts forward physically and technologically stringent requirements for its central detector (CD), including a large volume and target mass (20 kt liquid scintillator, LS), a high energy resolution (3$\%$ at 1 MeV), a high light transmittance, the largest possible photomultiplier (PMT) coverage, the lowest possible radioactive background, etc. The CD design, using a spherical acrylic vessel with a diameter of 35.4 m to contain the LS and a stainless steel structure to support the acrylic vessel and PMTs, was chosen and optimized. The acrylic vessel and the stainless steel structure will be immersed in pure water to shield the radioactive background and bear great buoyancy. The challenging requirements of the acrylic sphere have been achieved, such as a low intrinsic radioactivity and high transmittance of the manufactured acrylic panels, the tensile and compressive acrylic node design with embedded stainless steel pad, one-time polymerization for multiple bonding lines. Moreover, several technical challenges of the stainless steel structure have been solved: the production of low radioactivity stainless steel material, the deformation and precision control during production and assembly, the usage of high strength stainless steel rivet bolt and of high friction efficient linkage plate. Finally, the design of the ancillary equipment like the LS filling, overflowing and circulating system was done.
}


\begin{document}
\maketitle
\flushbottom

\section{JUNO Central Detector requirements}
\label{intro}

The JUNO experiment was proposed to determine the neutrino mass hierarchy by measuring with high precision the energy spectrum of reactor anti-neutrinos at an intermediate baseline \cite{RefLMAMSSNP, RefPNOPIBRNE, RefDNMHIB}. To realize this goal, JUNO’s neutrino target, the Central Detector (CD), was dimensioned with 20 kt of Liquid Scintillator (LS) and with the additional requirements of high energy resolution (3$\%$ or better at 1 MeV), and radioactive background from U and Th better than $ 10^{-15} $ g/g. After this initial proposal, the background requirements were made two orders of magnitude stricter in order for JUNO to be able to also study solar neutrinos in detail. For further information on the JUNO design and updated physics goals, refer to the JUNO conceptual design report \cite{RefJUNOCDR} and the updated publication on the detector and JUNO's physics goals \cite{RefNPwJUNO, RefJUNOPhysicsandDetector}.

The CD is a key component of the JUNO experiment to detect neutrinos coming from different sources (reactors, supernovae, atmosphere, the Earth and the Sun, etc). The whole CD structure must be stable and work in a reliable way for 20 years of operation, the designed lifetime of JUNO. A spherical shape LS vessel, which has the minimum surface area for the same volume, is the key design of the CD to optimize the number and cost of the photomultipliers (PMTs) but providing a very high detector surface coverage. The vessel needs to be transparent for scintillation light to pass through and be detected by the PMTs. The 20 kt of LS needs a spherical vessel with a diameter of 35.4 m. To achieve the energy resolution (the photo-statistics is the main contribution, so we mainly consider it during the detector scheme design \cite{RefJUNOPhysicsandDetector}), the following are needed:
\begin{enumerate}
\item the huge CD volume requires a LS light attenuation length of at least 20 m;
\item the PMTs for the CD need to have the coverage at least 75$\%$;
\item the PMTs’ photon detection efficiency must be greater than 27$\%$.
\end{enumerate}

The CD will be immersed into pure water to shield the LS from the radioactive background coming from the surrounding environment. A veto detector system (VETO) will be placed outside the CD to detect the cosmic ray background; the system is made of two sub-systems \cite{RefJUNOPhysicsandDetector}: a Water Cherenkov Detector and the Top Tracker \cite{RefJUNOETT}. To detect low energy neutrino, the CD must be made of low radioactive materials \cite{RefJUNORCSD}. All the parts in contact with the LS need to be compatible with the LS and properly cleaned to lower the background. The air-tightness of CD should be low enough to prevent possible Radon leakage into the LS.

\section{JUNO Central Detector Design Schemes}
In the beginning, several CD design schemes were proposed, with the following main requirements:
\begin{itemize}
\item design a structure capable of containing 20 kt of LS;
\item perform an accurate choice of material in order to minimize the radioactive background;
\item install a large number of PMTs outside CD;
\item realize a support structure ensuring a stability and reliability over the JUNO lifetime.
\end{itemize}

Four options where selected as design candidates. They are discussed in the following, taking into account not only the physics requirements but also the engineering feasibility, reliability, lifetime, and costs.

\subsection{Option 1: Acrylic sphere and steel structure}

This option, initially proposed in 2012, used a spherical acrylic vessel with a diameter of 35.4 m to contain the LS and a stainless steel structure to support the PMTs. In this design, the acrylic vessel and the stainless steel structure are immersed in pure water which shields the LS from the radioactive background coming from the experimental hall rocks, steel structure and glass of PMTs. The water also works as Cherenkov medium to detect cosmic muons. The inward-facing PMTs detect scintillation photons from the LS and the outward-facing PMTs detect the Cherenkov photons produced by the interaction of cosmic muons in the surrounding water. Both sets of PMTs are installed on the stainless steel structure.

The reasons of choosing acrylic as a container of the LS are the following:
\begin{itemize}
\item acrylic is fully compatible with LS and pure water;
\item acrylic is highly transparent for the light of the wavelength from 380 to 700 nm \cite{RefMSATJUNOCD, RefSATJUNOCD};
\item acrylic, i.e. polymethyl methacrylate (PMMA), is composed of atoms of Carbon, Hydrogen, and Oxygen, which contains very low radioactive backgrounds contamination \cite{RefPAHPUTCM, RefSUThRRAST};
\item acrylic has good mechanical properties, such as tensile strength, compressive strength, and resistance to creep;
\item acrylic has a long lifetime, and it has been used with success by several past neutrino experiments (SNO, SNO+, and Daya Bay) \cite{RefNIM449172, RefNIM03943}. Moreover it is largely employed in the construction of large size aquariums.
\end{itemize}

\begin{figure}[!ht]
    \centering
  \includegraphics[width=0.98\linewidth]{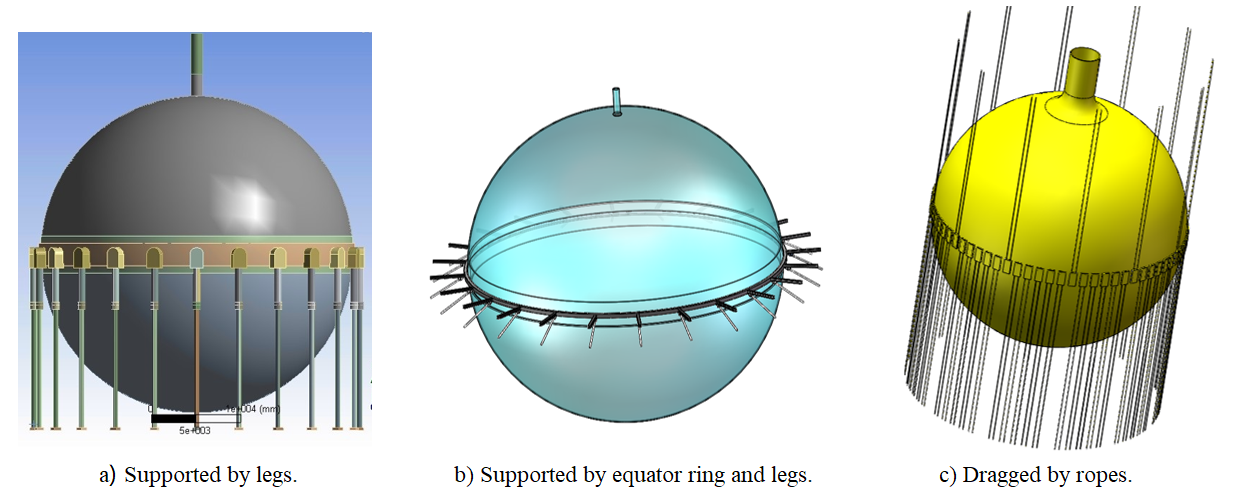}
  \caption{CD design, option 1: methods to support the acrylic sphere.}
\label{CDDesignOption1}       
\end{figure}

The huge acrylic spherical vessel will have a net weight of about 6x$10^6$ N. Since the LS density is 0.856 g/$cm^3$, the acrylic vessel will bear a buoyancy load of about 3x$10^7$ N once filled with 20 kt of LS and immersed in a pure water pool. How to support it during both the construction phase (i.e. no liquid inside) and running phase (i.e. LS inside and pure water outside) is a very big technical challenge. Several support methods have been proposed, designed, and analyzed. The first one is to use several support legs made of acrylic and stainless steel (Fig. \ref{CDDesignOption1}a). The second method is to support the acrylic sphere in the equator area using an embracing ring and stainless steel triangle legs (Fig. \ref{CDDesignOption1}b). The third method proposed to use high-strength ropes or straps to drag the acrylic sphere (Fig. \ref{CDDesignOption1}c).

\begin{figure}[!ht]
    \centering
  \includegraphics[width=0.98\linewidth]{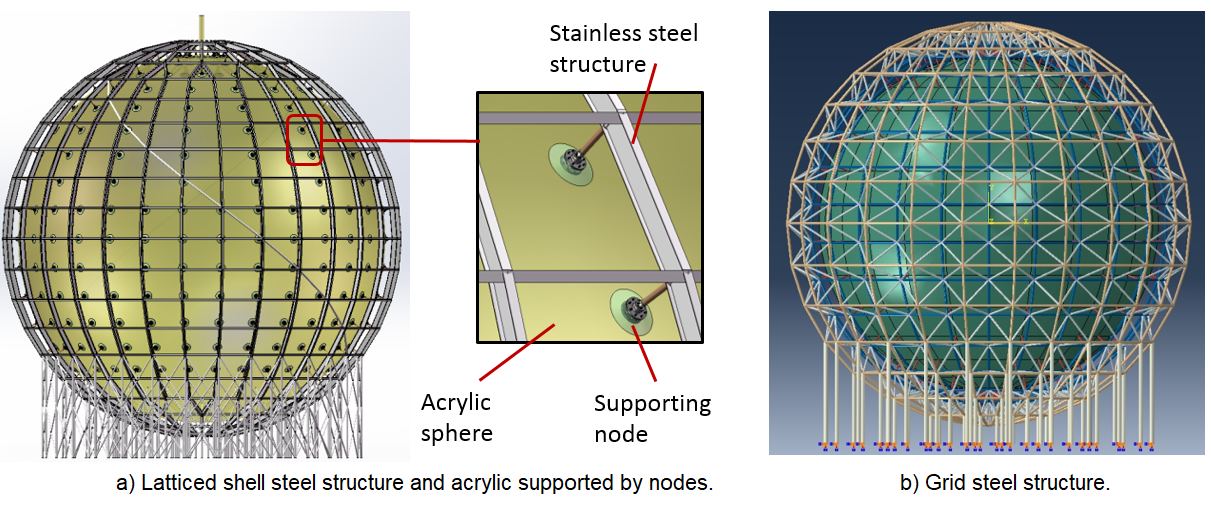}
\caption{Acrylic sphere supported on stainless steel structures through supporting nodes.}
\label{AcrylicSphere}       
\end{figure}

However, all the three methods mentioned above and shown in Fig. \ref{CDDesignOption1}  face similar issues: it is very difficult to reduce the stress of acrylic in the connection area and maintain a reasonable safety factor for such huge buoyancy. Finally, a fourth method was proposed and proved to be able to satisfy both requirements. It employed hundreds of acrylic nodes to support the acrylic sphere over the whole surface. The acrylic node (details in Section \ref{SectionAcrylicNode}) uses stainless steel pad structure embedded into an additional acrylic pad structure, which is bulk-polymerized onto the acrylic sphere. The fourth method was selected and two types of stainless steel pad structures were proposed and analyzed, as shown in Fig. \ref{AcrylicSphere} . One is called grid structure (Fig. \ref{AcrylicSphere}b) and the other latticed shell structure (Fig. \ref{AcrylicSphere}a). The shell structure was finally adopted because of easier PMTs arrangement and installation and also because it reduced the radial space occupied in the water pool to a minimum.

\subsection{Option 2: Acrylic boxes and steel tank}

\begin{figure}[!ht]
    \centering
  \includegraphics[width=0.98\linewidth]{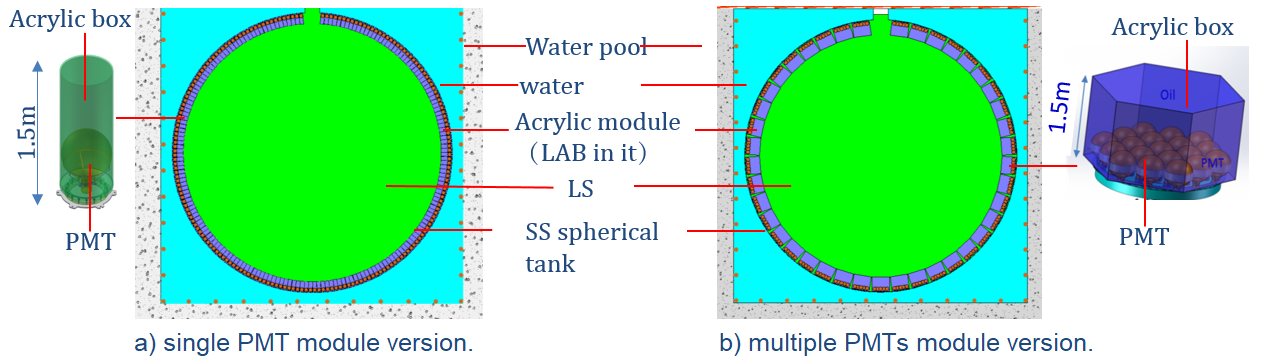}
\caption{Conceptual design of the Acrylic Boxes in a Steel Tank.}
\label{AcrylicBoxedSteelTank}       
\end{figure}

To reduce the difficulty and risk for building a huge acrylic spherical vessel, an option called acrylic boxes and stainless steel tank was proposed to split the acrylic sphere into many small acrylic modules and separate the LS in the CD and the LAB (Linear Alkyl Benzene) inside the modules. The conceptual design is shown in Fig. \ref{AcrylicBoxedSteelTank}: several acrylic modules are arranged and positioned in the inner surface of a stainless steel spherical tank. The interior of each acrylic module is filled with LAB and the LS is filled outside of the module, inside the stainless steel tank. The buffer between the stainless steel tank and the water pool is filled with pure water. Two type of modules are possible: a single PMT module (Fig. \ref{AcrylicBoxedSteelTank}a) where each module contains only one PMT, and a multi-PMTs module, where several PMTs are sealed in one bigger module (Fig. \ref{AcrylicBoxedSteelTank}b). The height of each acrylic module is 1.5 m.

\begin{figure}[!ht]
    \centering
  \includegraphics[width=0.8\linewidth]{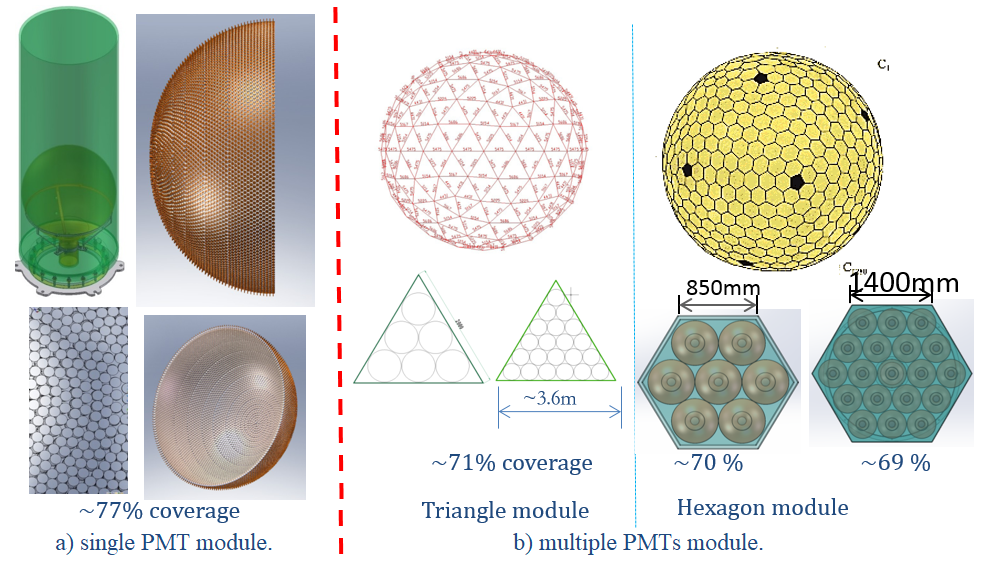}
\caption{\centering PMT coverage for two kinds of PMT module.}
\label{PMTcoveragefortwo}       
\end{figure}

Fig. \ref{PMTcoveragefortwo} compares the PMT coverage of two kinds of modules. The multiple PMTs module, either with hexagon or triangle arrangement (Fig. \ref{PMTcoveragefortwo}b), has a lower coverage with respect to the single PMT module (Fig. \ref{PMTcoveragefortwo}a). Moreover, since the single PMT module is smaller and lighter, the operations of LAB filling inside each module, sealing, and leak checking are easier to be performed before installing the modules in the stainiless steel tank. Therefore, this configuration was selected for further design and tests.

\begin{figure}[!ht]
    \centering
  \includegraphics[width=0.98\linewidth]{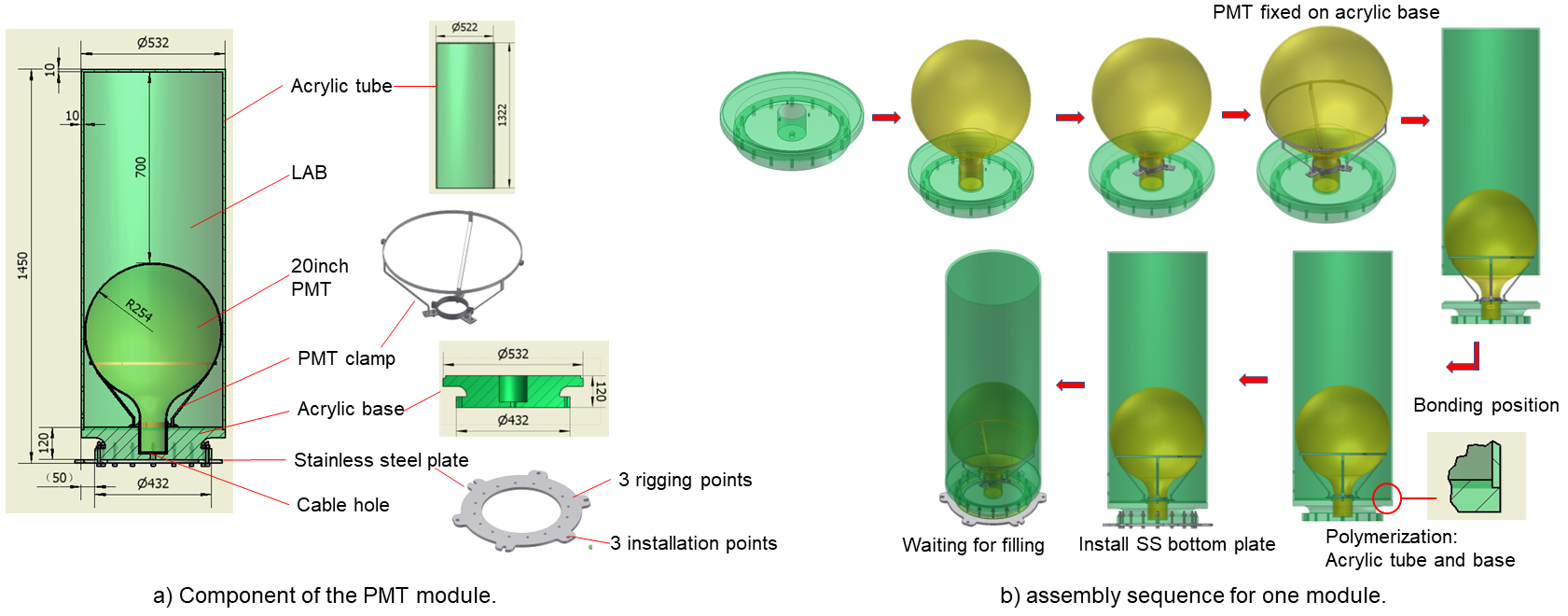}
\caption{Single PMT module and the assembly sequence.}
\label{SinglePMTmodule}       
\end{figure}

Fig. \ref{SinglePMTmodule} shows the components of the single PMT module and its assembly sequence. The acrylic parts can be machined at the manufacturing company and the module assembly can be done at the experimental site. The acrylic tube needs to be polymerized with the acrylic base to ensure the module performs reliably after the detector construction for the duration of the experiment. LAB is filled into the module after assembly and a leak check is performed before installation in the Steel Tank. Only one cable hole (for PMT readout) is available at the bottom of the module, and therefore a leak check needs to be done after the acrylic body is polymerized together.

\subsection{Option 3: Balloon and steel tank}

\begin{figure}[!ht]
    \centering
  \includegraphics{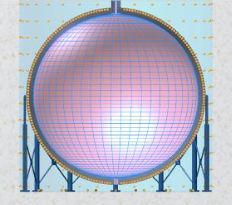}
\caption{Schematic view of the option of balloon and stainless steel tank.}
\label{BalloonandStainlessSteelTank}       
\end{figure}

The third option for the CD was designed with an inner balloon and an outer stainless steel tank, following the design of previous liquid scintillator neutrino experiments, such as KamLAND \cite{RefKamLAND2006} and Borexino \cite{RefBorexinoDetector}. In the design, 20 kt of LS are contained in a 35 m in diameter membrane sphere which can be made of transparent nylon 6 film with a thickness between 0.1 mm and 0.2 mm. On the outside of the balloon, a stainless steel tank with a diameter of about 38 m is set up in the center of a 44 m deep water pool. The buffer liquid filled between the tank and the balloon can be either LAB or mineral oil, whose density is 0.3–0.5$\%$ smaller than that of the LS. The balloon will bear the pressure load from the density difference between the inner and outer liquid density. About 18,000 20-inch PMTs can be fixed to the inner surface of the stainless steel tank. A liquid filling pipe is connected to the top of the balloon by a flange. Fig. \ref{BalloonandStainlessSteelTank} shows the structural sketch of this option.

\begin{figure}[!ht]
    \centering
  \includegraphics{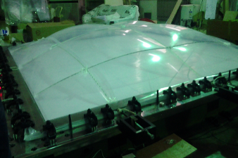}
\caption{The experiment to test the local sphere membrane.}
\label{TestLocalSphereMembrane}       
\end{figure}

The balloon will be restrained and held by a set of ropes to bear the load of weight and buoyancy. The rope with high strength and low elongation and a diameter of 6–8 mm can be made of high-density polyethylene (HDPE). An experiment with a smaller section of the sphere membrane was conducted to test this option (Fig. \ref{TestLocalSphereMembrane}). The stresses and deformations of the membrane and ropes were measured at different pressure after having replaced different membranes and ropes and changing the mesh size of the rope. A Finite Element Analysis (FEA) was verified by the experimental data, and afterwards the entire balloon design was calculated. With a 0.3$\%$ density difference between the LS and the buffer liquid, the FEA result shows that the stress of the membrane can be less than 5 MPa and the deformation of the balloon less than 0.5 m by selecting suitable membrane and ropes. As an example, for a membrane with a thickness less than 0.2 mm and elastic modulus greater than 150 MPa, more than 96 ropes with a diameter greater than 6 mm, and elastic modulus greater than 70 GPa are required.

\begin{figure}[!ht]
    \centering
  \includegraphics{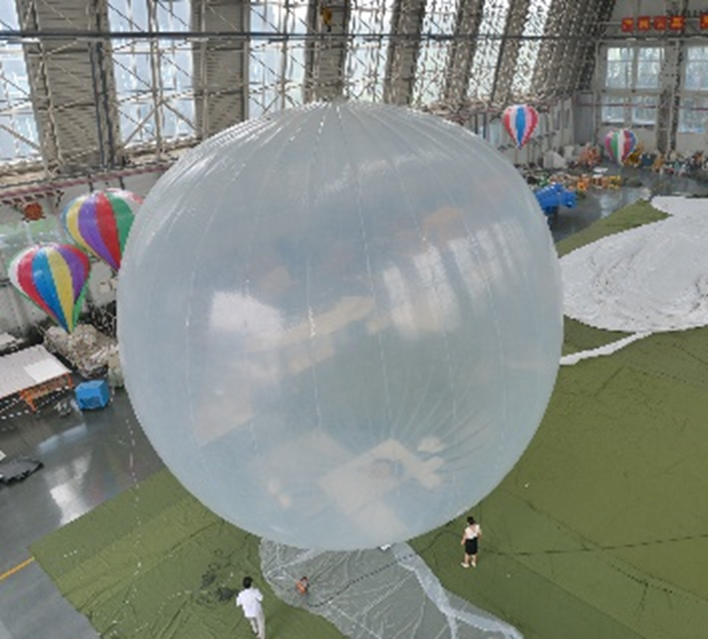}
\caption{A 12 m diameter balloon prototype.}
\label{BalloonPrototype}       
\end{figure}

A 12 m diameter balloon prototype was manufactured with nylon 6 film (0.127 mm thickness), as shown in Fig. \ref{BalloonPrototype}. It was a beneficial R$\&$D attempt to examine the production techniques and discover potential problems, which could put in danger future production for liquid scintillator neutrino experiments.

Preliminary studies showed that this option should be able to meet the requirements set by JUNO with the proper selection of materials and an optimization of the structural design. However, the biggest risk and challenge for this option is to completely avoid cracks or leaks throughout the entire life of the experiment.

\subsection{Option 4: Acrylic tank and steel tank}

\begin{figure}[!ht]
    \centering
  \includegraphics[width=0.98\linewidth]{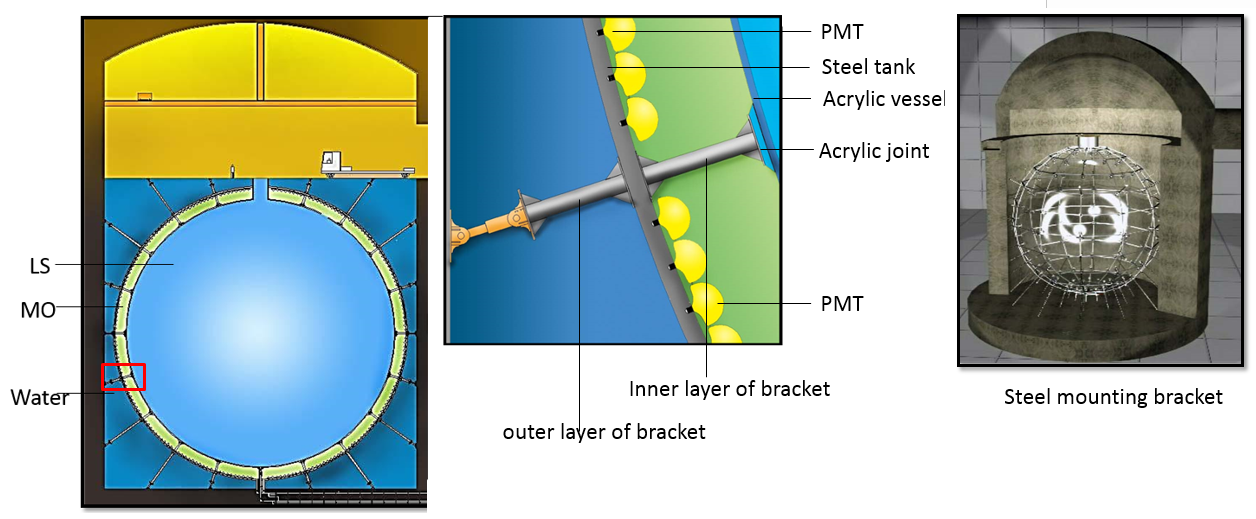}
\caption{Conceptual design of acrylic vessel and steel tank.}
\label{AcrylicVesselandSteelTank}       
\end{figure}

An alternative option for the CD design is an acrylic vessel inside a stainless steel tank as shown in Fig. \ref{AcrylicVesselandSteelTank}. This option was proposed to reduce the stress on the acrylic vessel resulting by filling mineral oil (MO) between the vessel and the tank. The stainless steel structure is made of two layers of stainless steel mounting brackets. The inner layer is designed to support and fix the acrylic spherical vessel and the outer layer is designed to fix the whole detector to the concrete wall of the water pool. For the realization of the design, 132 fixing bracket points are required. 

\begin{figure}[!ht]
    \centering
  \includegraphics[scale=0.3]{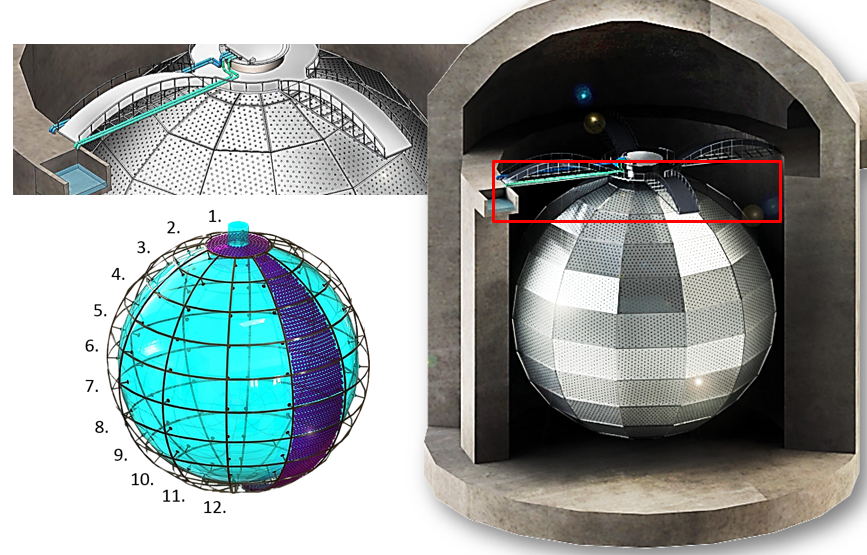}
\caption{PMT arrangement and steel tank.}
\label{PMTArrangementandSteelTank}       
\end{figure}

The PMTs will be divided into 132 groups along longitude and latitude, as shown in Fig. \ref{PMTArrangementandSteelTank}. The PMTs in each group will be mounted on stainless steel rear panels, and all of the panels will be welded to the sealed stainless steel tank. The detector will have two layers: the inner vessel made of acrylic and filled with 20 kt of LS, and an outer layer made by the stainless steel tank, filled with MO. Since LS and MO have a very similar density, the pressure inside and outside the acrylic vessel is balanced and the overall stress on the acrylic sphere is largely reduced with respect to other options. However, this option suffers from important limitations and possible risks coming from building the two vessels underground, in a limited space. Moreover, additional risks can arise from the steel tank manufacturing since the welding of its pieces underground can only be done once the acrylic vessel is already in place.

\subsection{Final design option determination}

Option 2, made of acrylic boxes and a steel tank, was not adopted mainly because of the risk of having residual LS remains in the gaps between PMT modules: LS photons will be produced due to the radioactive background from nearby materials like the PMT glass and steel and constitute a background to neutrino detection.

Option 3, namely a balloon and steel tank, was not adopted because of several reasons:
\begin{itemize}
\item a huge balloon may produce irreparable defects during production, transportation, and installation. There would be a risk of cracks or leaks as it would be subjected to significant stresses. Several small prototypes of balloons have shown such risks;
\item the balloon is prone to generate static electricity, so it is difficult to prevent dust from settling and to clean it during production and installation;
\item the shielding liquid of LAB or mineral oil will emit a small amount of light as particles pass through it;
\item while the balloon itself is not expensive, this option requires a shield of approximately 9.5 kt of LAB or mineral oil outside the balloon, and a stainless steel spherical container with a diameter of 40 m. Both of these additions bring additional costs to the detector construction.
\end{itemize}

Option 4, acrylic vessel inside a stainless steel tank, was abandoned because of two reasons:
\begin{itemize}
\item it is very difficult to construct two spherical vessels, one made of acrylic and the other stainless steel, in the water pool;
\item it is more expensive because of the shielding material of MO and two vessels. 
\end{itemize}

Option 1, i.e. an acrylic sphere and a steel structure, was ultimately chosen by the JUNO collaboration in 2015. Its optimization, engineering design, and new technologies are introduced in the following sections. Table \ref{ComparisonsCDdesign} reports a comparison of the different options with respect to a few design parameters.

\begin{table}[!ht]
\caption{Comparisons of four options of CD design}
\label{ComparisonsCDdesign}
\begin{tabular*}{\textwidth}{@{\extracolsep{\fill}}llllll@{}}
\hline
Items & \multicolumn{1}{c}{Option 1} & \multicolumn{1}{c}{Option 2} & \multicolumn{1}{c}{Option 3} & \multicolumn{1}{c}{Option 4}  \\
\hline
Physics impact & ***** & *** & **** & *****  \\
Engineering implementation & ***** & ***** & ***** & ***  \\
Break or leakage risks & ***** & **** & *** & ***** \\
Costs & ***** & *** & **** & **  \\
\hline
\end{tabular*}
Note: more stars means it is a better option with respect to this parameter.
\end{table}

\section{JUNO CD structure design}

\subsection{Structure requirements}

The design of the JUNO CD needs to meet several requirements, both from Physics and Engineering; among which is the lifetime of the CD that must be 20 years. The JUNO detector will go through the following phases: construction, filling with LS and water, and running. The stress and deformation of the structure at each operation step needs to be analyzed to ensure its reliability. Moreover, the structure should be analyzed under the earthquake loading with a local seismic intensity of magnitude 7 \cite{RefSGMPZMC}. Finally, the effects of temperature variation must be also evaluated.

The stress on the acrylic vessel is of particular concern for the CD. For DayaBay \cite{RefAIDayaBayAD}, its acrylic vessel’s lifetime was designed to be 5 years and its stress limit was required to be less than 5.5 MPa. For SNO \cite{RefNIM449172}, its acrylic vessel’s lifetime was designed to be 10 years and its stress was required to be less than 4 MPa. Considering the JUNO detector will operate 20 years, the stress of its acrylic vessel is required to be less than 3.5 MPa. The acrylic vessel is supported by the stainless steel structure through many connecting bars. The stress level of the acrylic is related to the axial force of the connecting bar and the structure of the acrylic node. The distribution of the connecting bar should be optimized to reduce the axial force, and also the structure and manufacturing process of the acrylic node should be optimized to minimize the stress on the acrylic vessel.

The CD average temperature while running was determined to be 21 °C to balance the need of heat flow for the front-end electronics, pure water, and electrical power. The average temperature while acrylic panels are machined in the workshop should be kept as the average temperature of the CD while running to control the variation of the dimensions of the acrylic vessel since the acrylic has a larger thermal expansion coefficient, 7 x $10^{-5}$ $K^{-1}$, compared to stainless steel. Moreover, the temperature for acrylic vessel construction in the underground hall should be the same to control the stress caused by the difference of thermal expansion coefficients of the acrylic and the stainless steel. 

The stainless steel structure bears almost all load and supports the acrylic vessel, the PMTs, the front-end electronics, the anti-geomagnetic field coils, and the panels optically separating the CD and the VETO. A total of 17,600 20-inch diameter and 25,600 3-inch diameter PMTs are placed facing the CD to detect scintillation light and 2,400 20-inch diameter PMTs are placed facing the opposite direction of the CD to detect Cherenkov light in the VETO. To obtain a CD PMT coverage of over 75$\%$, it is required that the gaps between PMTs should be as small as 3 mm. Therefore, the stainless steel structure should have precise dimensions and small deformation during fabrication, installation, and operation.

Referring to the design criteria of the steel structure, the nonlinear stability of the CD must have a safety factor of 2.5 or better \cite{RefDRMSSSCDofJUNO}. The design of the detector also must take into account the feasibility and costs of the assembly and installation.

\subsection{Structure optimization}

For the chosen structure, the inner diameter of the acrylic vessel is 35.4 m with 120 mm thickness; while the stainless steel is a frame structure made of H-beam with an inner diameter of 40.1 m. The acrylic vessel will be supported on the stainless steel structure by connecting bars. To reduce the internal forces on the connecting bars and the stresses on the acrylic and improve the safety of the detector structure, a series of factors were studied and optimized. A prototype of the CD was made and the design was tested in 2018 \cite{RefDSPCDofJUNO}.

\subsubsection{Optimization of the stainless steel latticed shell}

\begin{figure}[!ht]
    \centering
  \includegraphics[width=0.98\linewidth]{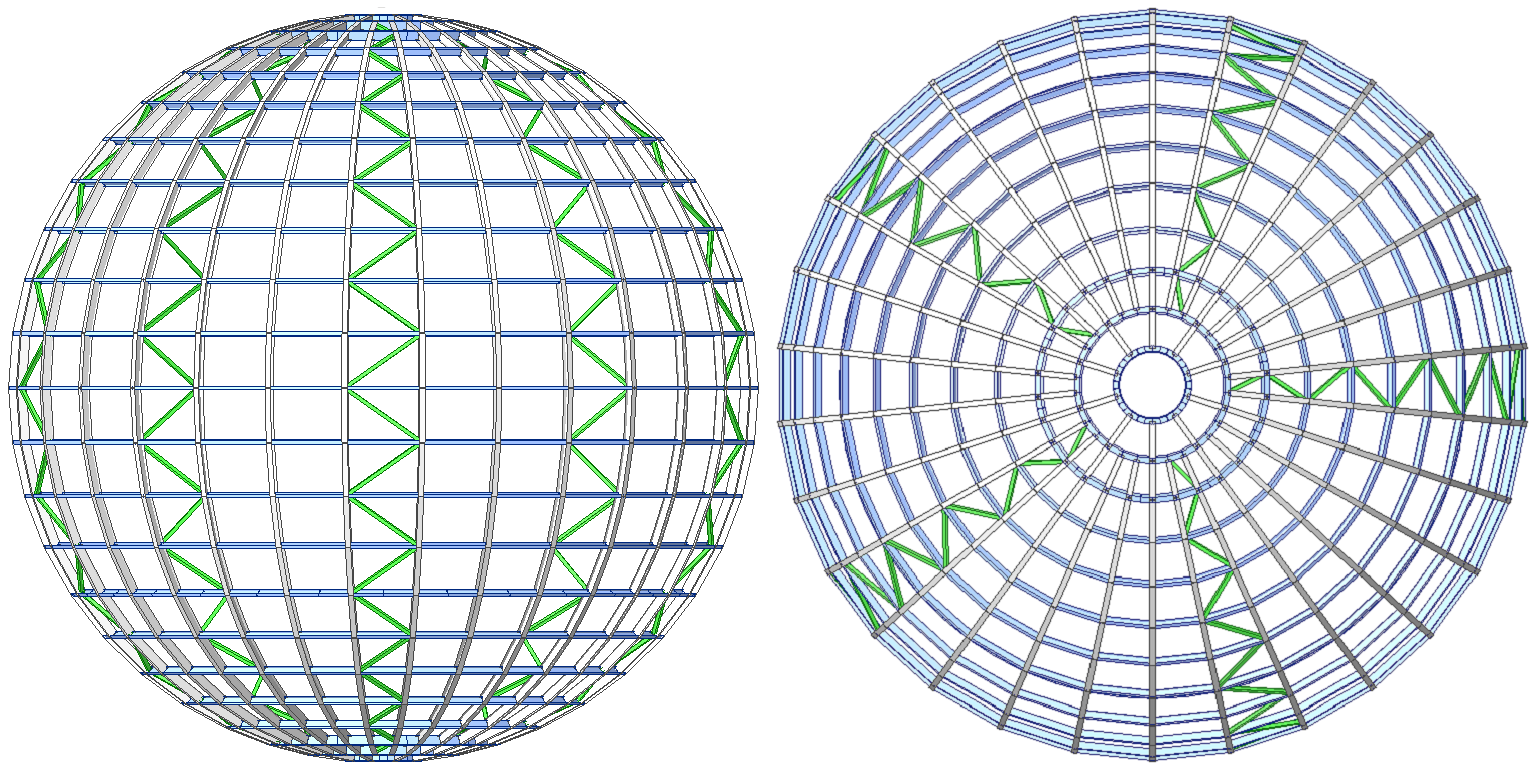}
\caption{Stainless steel latticed shell.}
\label{StainlessSteelLatticedShell}       
\end{figure}

Various reticulated shells, such as triangle and quadrilateral, were compared. The conclusions are that the quadrilateral frame has the advantage of a better control of the mechanical constraints for a controlled cost \cite{RefDSPCDofJUNO}. Taking into account the size of the acrylic panel, the light-blocking caused by the connecting nodes, the total length of the bonding lines, the axial forces, and few other factors, the stainless steel structure was finally determined with the format of the latticed shell composed of 23 latitudinal zones and 30 longitudinal zones, and the upper and lower hemispheres are symmetrically arranged, as shown in Fig. \ref{StainlessSteelLatticedShell}. To improve the torsional stiffness and stability of the structure, five in-plane zig-zag supports are evenly arranged along the meridional direction (see Fig. \ref{StainlessSteelLatticedShell})

\subsubsection{Optimization of the stainless steel supporting structure}

\begin{figure}[!ht]
    \centering
  \includegraphics[width=0.98\linewidth]{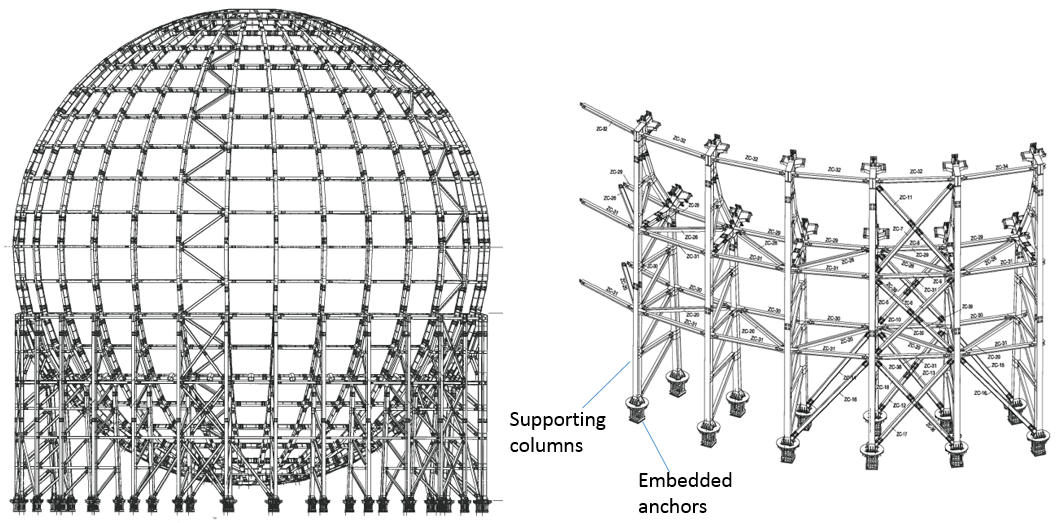}
\caption{CD supporting structure.}
\label{CDSupportingStructure}       
\end{figure}

The supporting structure of the CD is at the bottom, and is mainly composed of thirty plane truss columns, two circumferential horizontal supports, and five circumferential cross supports, as shown in Fig. \ref{CDSupportingStructure}. As a result of the FEA calculations, the column is arranged near the equator since it is more conducive than the lower pole area to reduce the force of the connecting bars in the lower hemisphere. 

The supporting columns of the detector will be connected with 60 embedded anchors in two circles at the bottom of the water pool. By optimizing the geometric shape of the plane truss column and the direction of the internal diagonal web members, the reaction force of the 60 supports are evenly distributed. During operation, when the temperature is stabilized, the reaction force of the supports can be controlled within 60 t, satisfying the civil construction requirements.

\subsubsection{Optimization of the connecting structure between the acrylic vessel and the stainless steel structure}

\begin{figure}[!ht]
    \centering
  \includegraphics[width=0.98\linewidth]{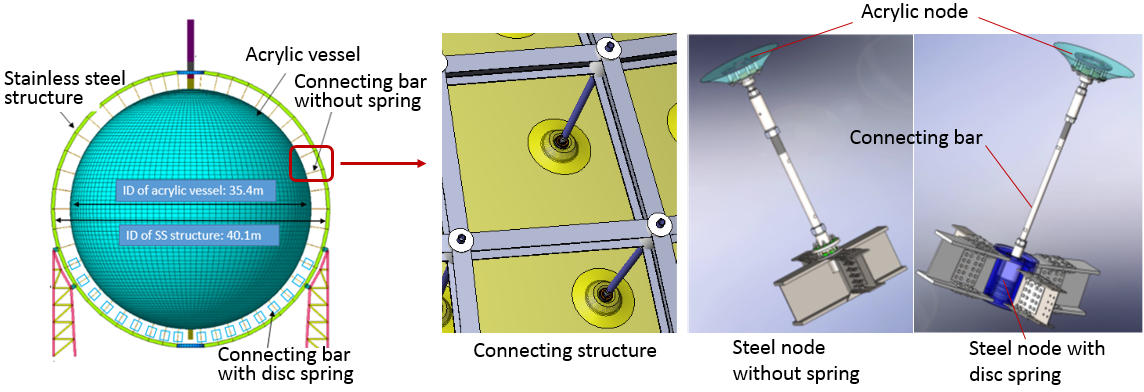}
\caption{Connecting structure between the acrylic vessel and stainless steel structure.}
\label{ConnectingStructure}       
\end{figure}

The connecting structure directly affects the stress level of the acrylic. The structure connecting the acrylic vessel and the stainless steel structure includes the acrylic node on the acrylic surface, the steel joint on the steel side, and the adjustable connecting bars. There are 590 connecting structures supporting the acrylic vessel on the steel structure outside the acrylic vessel. Both the acrylic node and steel node were designed to be hinged to realize the rotation in a small angle range, which is useful for installation. To prevent the potential buckling of the structure during detector running, a locking structure was added for 370 steel structure nodes which will be locked after installation. Moreover, a disc spring was used to adjust the axial stiffness and control the internal force for the other 220 steel nodes above the equator, as shown in Fig. \ref{ConnectingStructure}.

To monitor the forces on the connecting bars during installation, detector filling, and running, dedicated sensors will be installed on the bars and the force will be measured. Several measurement schemes were tested and compared, and measurement scheme by the 4 sensors of fiber bragg grating was finally chosen to monitor the status of the forces on the bars, achieving a measurement uncertainty of less than 0.7 kN \cite{RefRMCBAFofJUNO}.

\begin{figure}[!ht]
    \centering
  \includegraphics[width=0.98\linewidth]{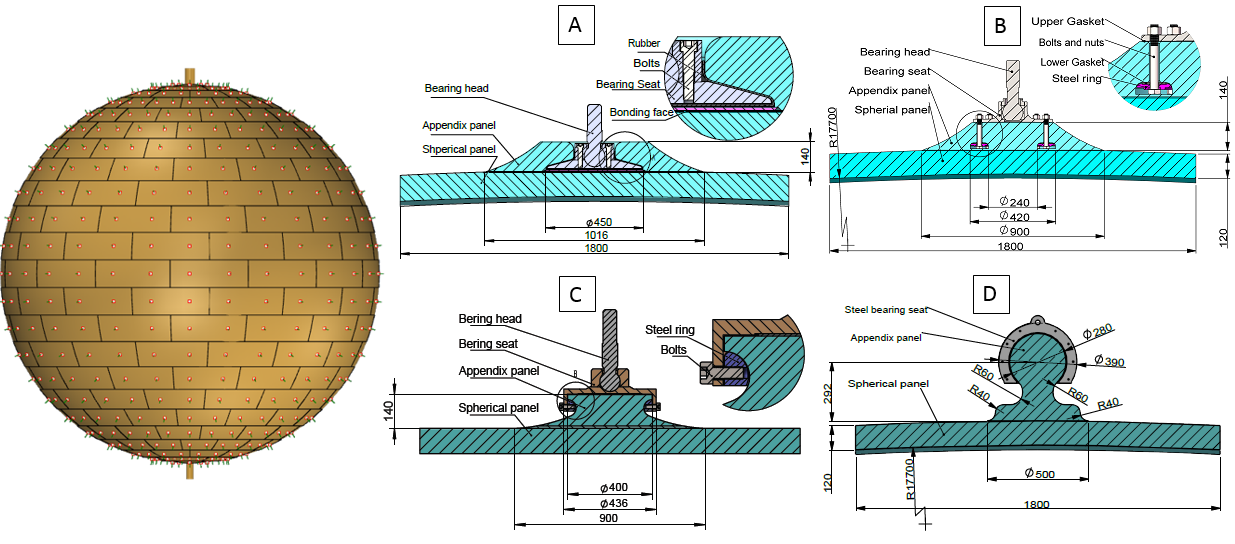}
\caption{Acrylic supporting nodes. The node type is noted next to the corresponding layout.}
\label{AcrylicSupportingNodes}       
\end{figure}

Besides the internal forces of the connecting bars, the acrylic node structure is very sensitive to the stress level of acrylic. Fig. \ref{AcrylicSupportingNodes} shows the layout of the 590 supporting nodes on the acrylic vessel and the 4 types of node structures, which were proposed at the beginning. Their comparisons are shown in Table 2. The acrylic stress of type B can be less than 3.5 MPa and its bearing capacity can be over 900 kN when tested by the prototype. Thanks to its outstanding performances, type B node was finally chosen as the final design \cite{RefSDCESNforJUNO, RefDEHLBACNLASV}.

\begin{table}[!htb]
\caption{Performance comparison of different node types. The first column indicates that the node is in a tensile and compressive load state. Under these two loads, the stress listed here is the maximum principal stress of the node.}
\label{PerformanceComparisonNodesTypes}
\begin{center}
\begin{tabular}{ |c|c|c|c|c| } 
\hline
\multicolumn{2}{| c |}{Node Type} & Light block /$\%$ & Stress /MPa
& Ultimate force/KN  \\
\multicolumn{2}{| c |}{ } & & by FEA & by measurement \\
\hline
\multirow{2}{1em}{A} & Tension & \multirow{2}{2em}{2.35} & 3.91 & Max. 510 \\ & Compression & & 5.98 & — \\ 
\hline
\multirow{2}{1em}{B} & Tension & \multirow{2}{2em}{1.77} & 2.92 & \textgreater 900 \\ 
& Compression & & 3.00 & \textgreater 900 \\ 
\hline
\multirow{2}{1em}{C} & Tension & \multirow{2}{2em}{2.21} & 6.53 & — \\ 
& Compression & & 5.36 & — \\ 
\hline
\multirow{2}{1em}{D} & Tension & \multirow{2}{2em}{0.91} & 4.99 & — \\ 
& Compression & & 4.43 & — \\ 
\hline
\end{tabular}
\end{center}
\end{table}

\subsubsection{Optimization of the chimney}
\label{OptimizationoftheChimney}

\begin{figure}[!ht]
    \centering
  \includegraphics[width=0.98\linewidth]{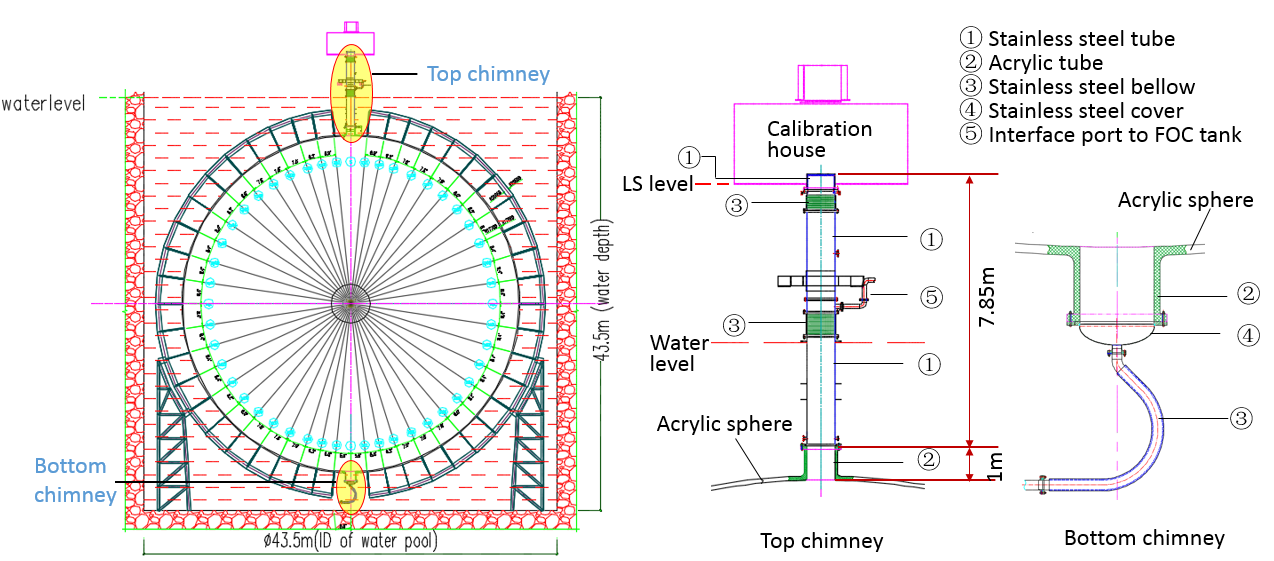}
\caption{Chimneys for the central detector.}
\label{ChimneysforCentralDetector}       
\end{figure}

There are two chimneys designed at the top and bottom of the CD acrylic sphere, as shown in Fig. \ref{ChimneysforCentralDetector}. The top chimney is the access port for the calibration operations during detector running and the rigging cable of some acrylic panels during installation; it will also be the main connection to the Filling, Overflowing and Circulating (FOC) apparatus, which is described in the following Section 7. During detector filling, the LS will enter near the top of the chimney keeping it 4.5 m higher than the water level outside the acrylic sphere, in order to improve the pressure distribution on the acrylic surface and optimize the force distribution, reducing the internal forces of the connecting bars and minimizing the maximum stress nearby on the acrylic surface. The inner diameter of the top chimney is 800 mm and the total length is 8.85 m: 1.00 m long in the acrylic section, and 7.85 m in the stainless steel section. The top chimney will be fixed to the top tracker bridge and enter the calibration house. To avoid extra load during transfer and absorb the deflection between CD and the other systems during installation, detector running, and in case of earthquakes, the welded bellows on the chimney have been done in two pieces. Between each tube of the chimney, there are the flanges and double o-rings bolted to ensure the sealing of the chimney. The O-rings are made of Viton-A material, and testing has proven their compatibility with liquid scintillator and pure water. Additionally, we have manufactured some samples of double o-rings sealed structures and conducted accelerated aging tests, demonstrating that there will be no leakage of liquid scintillator or water for 30 years. The bottom chimney is designed to provide access inside CD during cleaning of the acrylic vessel; moreover it can provide access to humans in case of dedicated special operations. A bent bellow on the bottom chimney is designed to absorb the displacement caused by the buoyancy of the acrylic sphere after filling, and will be connected to the acrylic sphere through a stainless steel cover flange. 

\section{Acrylic sphere R$\&$D}

\subsection{Research on acrylic performances}

\subsubsection{Mechanical performances}

The chemical formula for the JUNO acrylic is unique since, in contrary to commercially available acrylic panels, it does not contain any anti-UV material and plasticizer. Therefore its mechanical properties needed to be studied in detail, especially for the long-term operation performances. The tensile strength of 120 mm one-time cast panel can reach 70 ± 4 MPa. The elastic modulus can reach 3.1 GPa. The elongation at the break of an acrylic panel without the plasticizer will decrease from (4.9 ± 0.6)$\%$ to (3.8 ± 0.6)$\%$. The bending strength of one panel can reach 110 MPa. After optimization of the bonding process, the strength of the bonding area can reach more than 90$\%$ of the bulk panel.

In addition, the creep property of acrylic is critical to the design and long-term use in JUNO. The material’s long-term performance is related to its environment. The acrylic vessel will contain LS, placed in ultrapure water. The results of the tests show that the creep and aging performances of acrylic are worse in LS than in water. Therefore, long-term performance studies in LS environment were performed.

A customized creep machine was designed and used for the creep tests, which simulates the actual usage environment of the JUNO acrylic, such as the long-term immersion of the acrylic in liquid scintillator and pure water at 21°C \cite{zhou2016creep}. It can soak the acrylic sample in LS at 21 °C and apply certain stresses from 14 MPa to 40 MPa. These higher stresses can shorten the test time, which is useful for extrapolating creep rupture time under working stress. Load, deformation, and time were recorded during the test process. The creep rupture time under working stress can be extrapolated by fitting the test data with higher stresses applied.

\begin{figure}[!ht]
    \centering
  \includegraphics[width=0.65\linewidth]{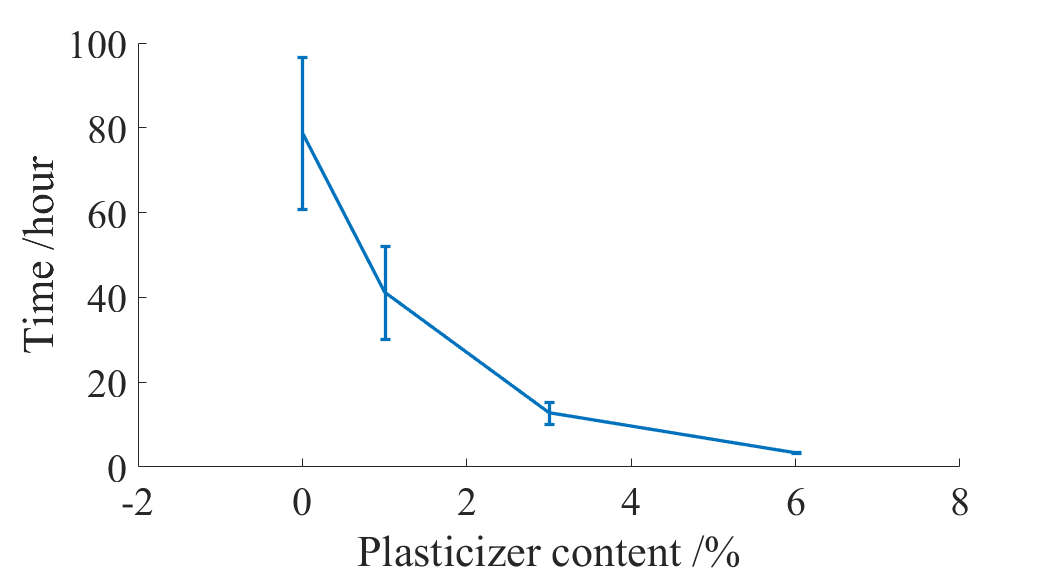}
\caption{Influence of plasticizer on creep. The Y axis is the creep fracture time.}
\label{InfluenceofPlasticizeronCreep}       
\end{figure}

The study looked at the creep performances of acrylic samples with different plasticizer content. According to Fig. \ref{InfluenceofPlasticizeronCreep}, the creep performances improve when the plasticizer content is lower. That is another reason why the acrylic formula of JUNO does not contain the plasticizer.

\begin{figure}[!ht]
    \centering
  \includegraphics[width=0.98\linewidth]{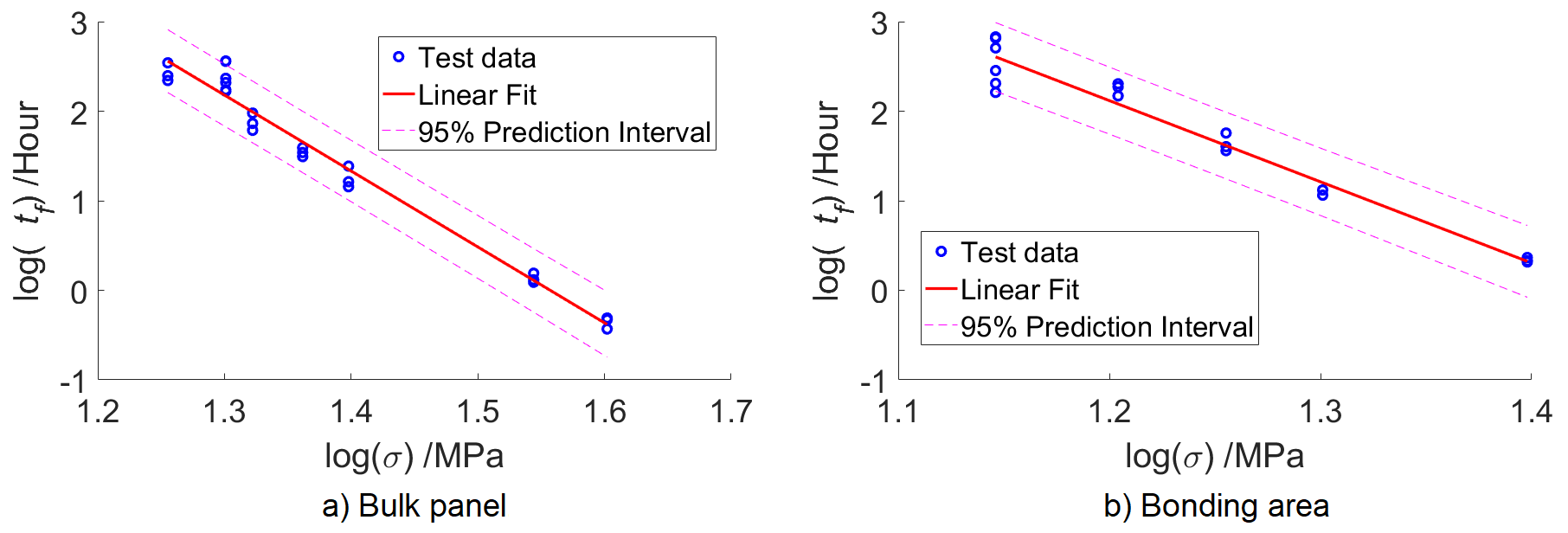}
\caption{Logarithmic graph of stress and fracture time.}
\label{LogarithmicGraphofStressandFractureTime}       
\end{figure}

The logarithm of the measured stress has a linear relationship with respect to the fracture time as shown in Fig. \ref{LogarithmicGraphofStressandFractureTime}. Through data fitting, the creep rupture time under a low-stress load of 3.5 MPa can be extrapolated according to the high-stress test data. The average creep rupture time for the bulk panes and bonding areas are 674 years and 104 years, respectively, resulting in a lower time limit of 230 years and 24 years (95$\%$ C.L.), respectively, as shown in the dashed lines in Fig. \ref{LogarithmicGraphofStressandFractureTime} .

\subsubsection{Optical performances}

Acrylic transmittance is an important optical performance characteristic of the CD and directly related to the number of detected photons, therefore, affecting the energy resolution. The transmittance requirement of the acrylic spherical panel of 120 mm thickness is of over 96$\%$ at 420 nm in ultrapure water, which is close to the peak wavelength of PMT quantum efficiency. A specific measurement system was developed to measure the transmittance of acrylic samples immersed in ultrapure water from 290 nm to 700 nm. 

Commercially available acrylic panels generally contain plasticizers and anti-UV agents. The formers usually improve acrylic machining performances, while the latters increase acrylic lifetime in environments with UV light. It was found that the plasticizers and anti-UV agents can reduce the acrylic transmittance, therefore they are not used in acrylic formula for the JUNO CD.

Thermoforming temperature and time were studied and optimized against acrylic transmittance. The roughness of the acrylic surface has an impact on the transmittance, diffusing reflection, and scattering of light. For this reason, grinding and polishing of the CD acrylic surface are mandatory to make it very smooth. Batch test results indicate that the average transmittance of mass-produced spherical acrylic panels is (96.5 ± 0.5)$\%$  at 420 nm and (95.5 ± 0.5)$\%$  at 413 nm \cite{RefMSATJUNOCD, RefSATJUNOCD}.

\subsubsection{Radioactivity control}

\begin{figure}[!ht]
    \centering
  \includegraphics[width=0.98\linewidth]{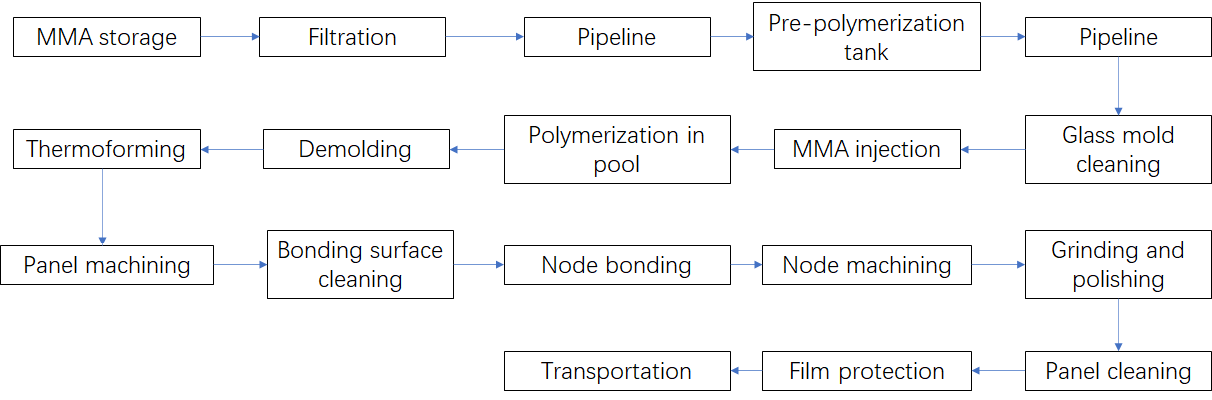}
\caption{Production process of acrylic panel.}
\label{ProductionProcessofAcrylicPanel}       
\end{figure}

The production process of the acrylic sheets is shown in Fig. \ref{ProductionProcessofAcrylicPanel}.

The radioactive background level of an acrylic panel directly affects measurement of the neutrino signal. The production line of acrylic at the Donchamp (Jiangsu) Materials Technology Co., ltd. \cite{RefDonchampMaterialsTechnology}, was specially developed for JUNO to meet our requirements. The following techniques were adopted to minimize radioactivity contamination during production:

\begin{enumerate}
\item The MMA (methyl methacrylate) raw material is filtered and purified to remove impurities.
\item Special pre-polymerization tanks and transportation pipes, dedicated to JUNO production, are cleaned by ultrapure water and MMA.
\item The glass mold is cleaned by ultrapure water in a cleaning room.
\item The acrylic node and bonding area parts are cleaned before the acrylic node being bonded to a spherical panel. A special equipment is used to inject MMA glue by tubes, avoiding MMA contact to environmental air.
\item The flat and spherical panels are protected with plastic films to avoid contamination from the daughters of radon.
\item Before the spherical acrylic panels are transported to the site, they are ground and polished to a depth greater than 0.1 mm and then the surface of the panel is cleaned and coated with a protective plastic film.
\end{enumerate}

The average \textsuperscript{238}U contamination in the JUNO spherical panel is 0.6 ppt, while that coming from \textsuperscript{232}Th is about 0.7 ppt; both are less than the required value of 1 ppt \cite{RefJUNORCSD}.

\subsection{Critical techniques of acrylic vessel}

\subsubsection{Thermoforming (molds, temperature field, shape survey results)}

Acrylic will soften and deform above glass-transition temperature. First, the flat panel is placed on a spherical mold and put into a baking room. The room temperature can be controlled according to a set temperature curve, which defines three separate stages: heating, heat preservation, and cooling. A certain external pressure is applied on the edges of the panel to ensure that the panel can fit well with the mold, thus obtaining a spherical shape panel with the required size. The pressure marks formed around the panel edges need to be machined off at a later stage. In the process of thermoforming, the surface and center temperatures of the panel should be consistent, in order to minimize the thermal stresses, the dimensional deviations, and rebound of the panels \cite{RefFEATBLSPMMAPinJUNO}.

The CD acrylic vessel diameter is required to be 35,400 ± 50 mm. For each panel, its thickness is required to be 124 ± 4 mm. The deviation between a spherical panel and a standard sphere is required to be less than 6 mm. The measurement results of the thermoformed panel in Fig. 19 show that the deviation of measured points changes regularly, which indicates that the panel has a specific shrinkage. However, the deviation between measuring points is less than 6 mm, satisfying the CD technical requirements.

\begin{figure}[!ht]
    \centering
  \includegraphics[width=0.65\linewidth]{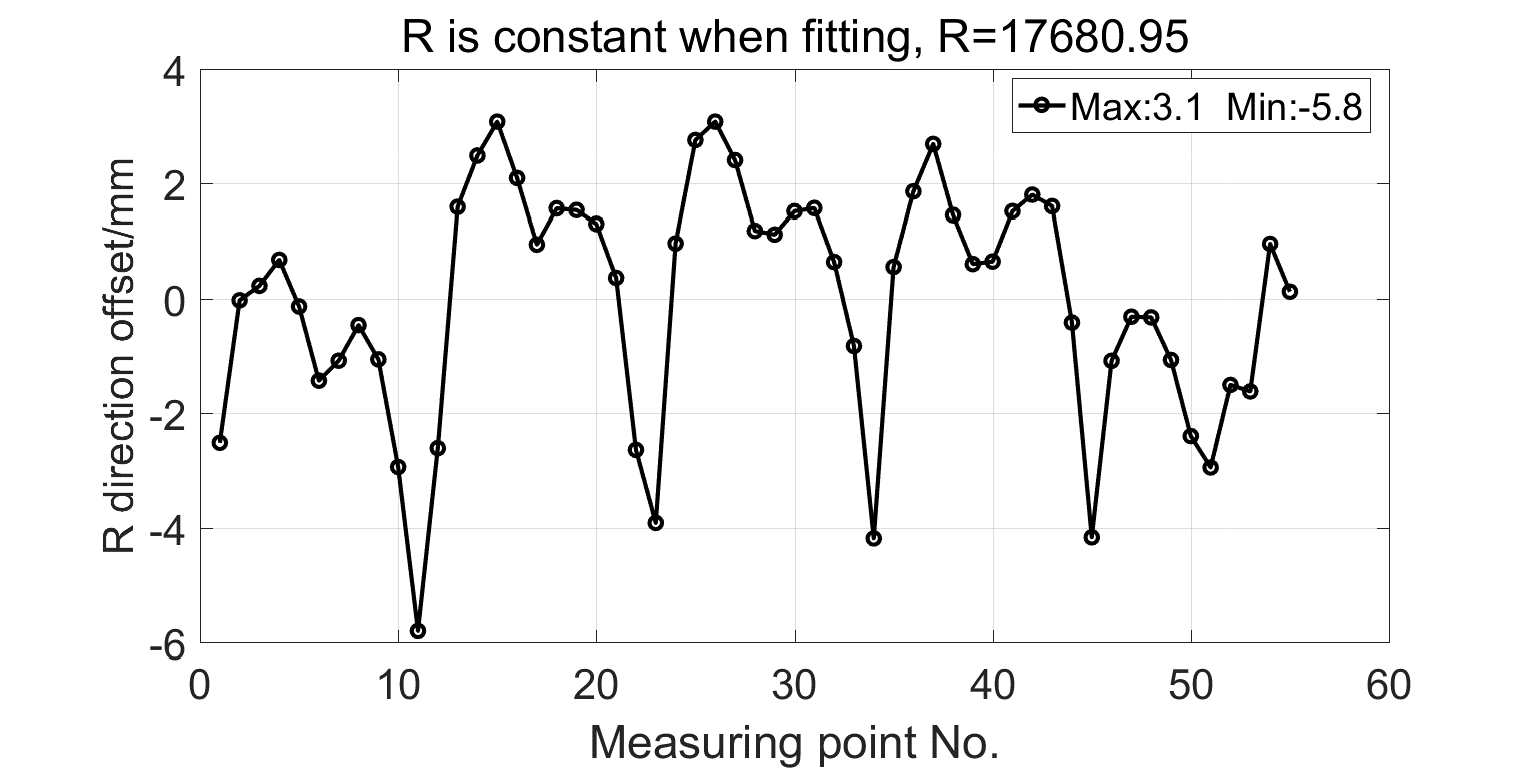}
\caption{Acrylic panel spherical shape measurements. The x-axis is the measuring point, with a total of 5 x 11 measuring points. The y-axis indicates the deviation between each point and the analytical sphere.}
\label{AcrylicPanelSphericalShapeMeasurements}       
\end{figure}

The properties of the thermoformed acrylic panel changed a little after production and became slightly brittle. The breaking strain decreased from (4.9 ± 0.6)$\%$ to (3.8 ± 0.6)$\%$. The average light transmittance of the spherical panel decreased from 98.5$\%$ to 96.5$\%$.

\subsubsection{Acrylic node}
\label{SectionAcrylicNode}

\begin{figure}[!ht]
    \centering
  \includegraphics[width=0.98\linewidth]{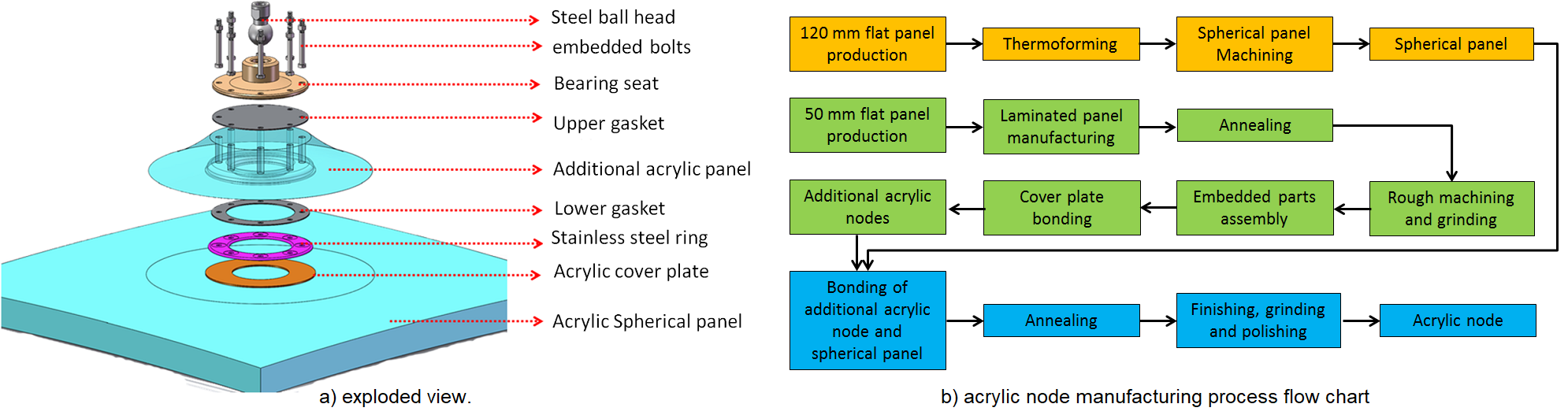}
\caption{Acrylic node.}
\label{AcrylicNode}       
\end{figure}

The acrylic node structure is shown in Fig. \ref{AcrylicNode}a. It consists of a spherical panel, some embedded parts (stainless steel ring, bolts, and e-PTFE gasket), an additional acrylic panel, and an acrylic cover plate. Its manufacturing includes three processes, as shown in Fig. \ref{AcrylicNode}b. At the beginning, the plane needs to be machined on the spherical panel for bonding with additional acrylic node assembly. Afterwards, the cleaned embedded parts are put into the ring groove of an additional acrylic panel, and then the ring groove is sealed using a fast glue with the acrylic cover plate. Finally, the acrylic node assembly is bonded onto the machined plane of the spherical panel. At the very end, annealing and secondary finishing of the acrylic node arc-surface are carried out.

\subsubsection{Bulk polymerization techniques (one-step vertical bonding, big circular bonding, annealing curve)}
\label{BulkPolymerizationTechniques}

\begin{figure}[!ht]
    \centering
  \includegraphics[width=0.4\linewidth]{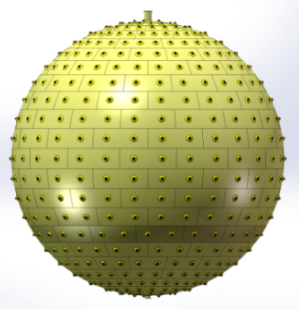}
\caption{PMMA spherical vessel segmentation scheme.}
\label{PMMASphericalVesselSegmentationScheme}       
\end{figure}

The PMMA spherical vessel is divided into 23 layers with 263 panel pieces and additional chimney structures at the two poles, as shown in Fig. \ref{PMMASphericalVesselSegmentationScheme}; the panels will be bonded layer by layer from top to bottom. The bonding process mainly includes three steps:

\begin{itemize}
\item sealing dams for containment of polymerizable cement is formed from acrylic strips and adhesive tapes;
\item cement is poured into the bond gaps and then polymerized for about 24 hours;
\item the bonding areas are heated for a certain time, during the annealing process, to further improve the polymerization process and obtain better performances.
\end{itemize}

\begin{figure}[!ht]
    \centering
  \includegraphics[width=0.98\linewidth]{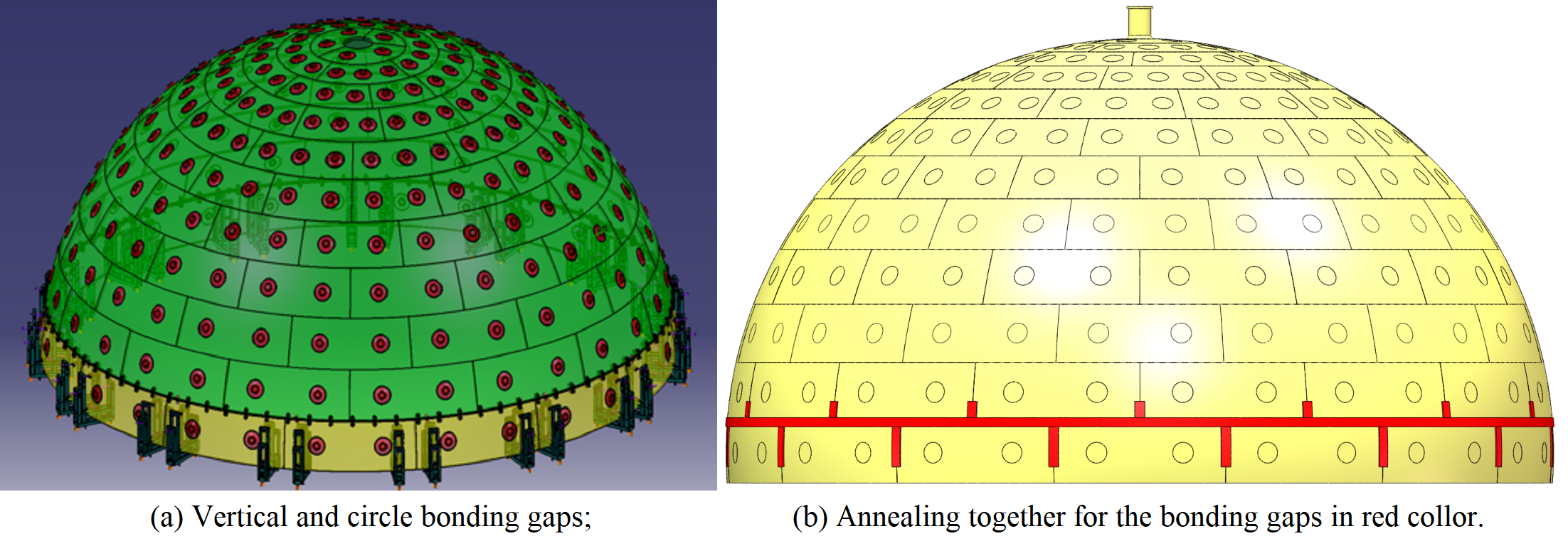}
\caption{Polymerization process of PMMA spherical vessel.}
\label{PolymerizationProcessofPMMASphericalVessel}       
\end{figure}

To optimize the construction time and positioning precision, all the operations (vertical bonding in one layer and circular bonding to the above layer) will be carried out at the same time. When the panels are located in a circle, the dams of all vertical bonding gaps and the circular bonding gaps will be formed and filled with cement, and then polymerized. The horizontal and vertical bonding gaps of a whole ring will be annealed together to reduce the risk of cracking caused by thermal stress and optimize the processing time. The polymerization process is shown in Fig. \ref{PolymerizationProcessofPMMASphericalVessel}.

\section{Stainless steel structure R$\&$D}

\subsection{Low radioactivity stainless steel}

The materials of the CD Stainless Steel (SS) structure need to be compatible with pure water, have low permeability and radioactive background contamination. The SS type 304, which is an austenitic SS according to ASTMA240/240M, was selected to build the CD structure. Its radioactive contamination limits to the expected event rate are reported in Table 3, based on the Monte Carlo simulations \cite{RefJUNORCSD}. The SS connecting bars and the SS parts embedded into the acrylic nodes are closer to the LS and their radioactive requirements are therefore much tighter.

\begin{table}[!ht]
\centering
\caption{Radioactive background requirement of SS.}
\label{RadioactiveBackgroundRequirementofSS}
\begin{tabularx}{0.98\textwidth}{@{\extracolsep{\fill}}|l|c|c|c|c|c|c|c|@{}}
\hline
Components & Mass & \textsuperscript{238}U & \textsuperscript{232}Th & \textsuperscript{40}K & \textsuperscript{60}Co & Singles & Upper\\
 & (T) & (ppb) & (ppb) & (ppb) & (mBq & in FV & limit\\
 &  &  &  &  & /kg) & (Hz) & (Hz)\\
\hline
Structure member & 583.79 & 1 & 3 & 0.2 & 20 & 0.02 & 0.1 \\
\hline
SS node & 23.6 & 0.2 & 0.6 & 0.02 & 1.5 & 0.9 & 1.5 \\
\hline
Connecting bar & 67.18 & 0.2 & 0.6 & 0.02 & 1.5 & 0.2 & 0.5 \\
\hline
Welding material & 13.35 & 200 & 100 & 10 & 50 & 0.02 & 0.05 \\
\hline
\end{tabularx}
Note: \textsuperscript{238}U: 1 ppb=12.4 mBq/kg; \textsuperscript{232}Th: 1 ppb=4.07 mBq/kg; \textsuperscript{40}K: 1 ppb=265 mBq/kg.
\end{table}

Low background SS materials and welding materials are respectively customized from Taiyuan Iron $\&$ steel Co., Ltd. \cite{RefTaiyuanIronSteel} and Tianjin Golden Bridge Welding Materials Group Co., Ltd. \cite{RefTianjinGoldenBridgeWeldingMaterialsGroup}. To achieve the above requirements, the following techniques were adopted:
\begin{itemize}
\item the Argon Oxygen Decarburization (AOD) process for ingot-refining hot metal from blast furnace was used during stainless steel production;
\item the smelting process used electrolytic Ni with a purity exceeding 99.9$\%$;
\item no scrap SS was added among the original materials.
\end{itemize}

\begin{figure}[!ht]
    \centering
    \includegraphics[width=0.98\linewidth]{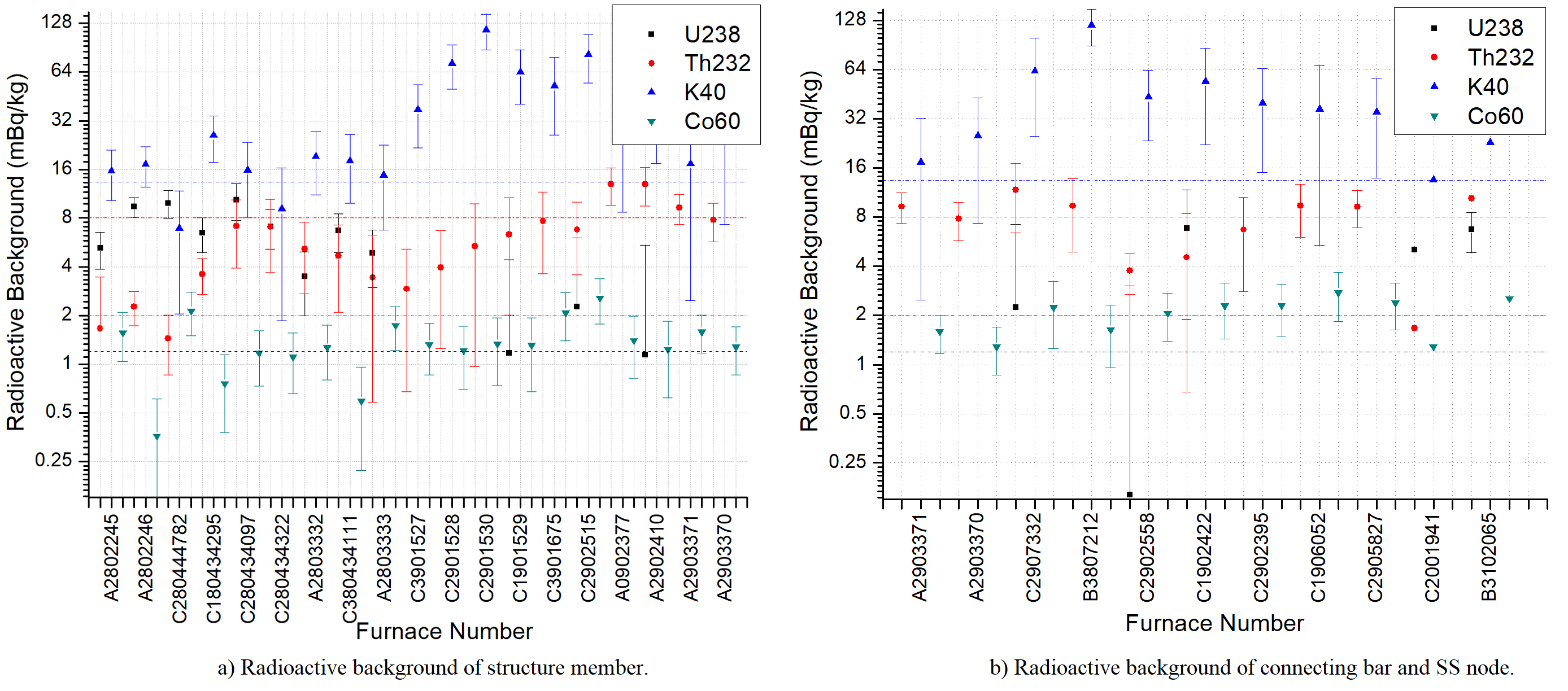}
    \caption{Stainless steel structure radioactivity contamination.}
    \label{SSStructureRadioactivityContamination}       
\end{figure}

In total, about 900 t of SS are needed for building the JUNO SS structure. Thirty four new furnaces for the SS production were specially realized for JUNO, and among them, thirty met the radioactive requirements of the experiment. The measurements results tested by China Jinping Underground Laboratory \cite{RefChinaJinpingUndergroundLaboratory} are shown in Fig. \ref{SSStructureRadioactivityContamination}. Although some measurements exceed the requirements, the overall radioactive background assessment meets the JUNO background goals \cite{RefJUNORCSD}.

\subsection{Deformation control: welding and preassembly}

\begin{figure}[!ht]
    \centering
    \includegraphics[width=0.6\linewidth]{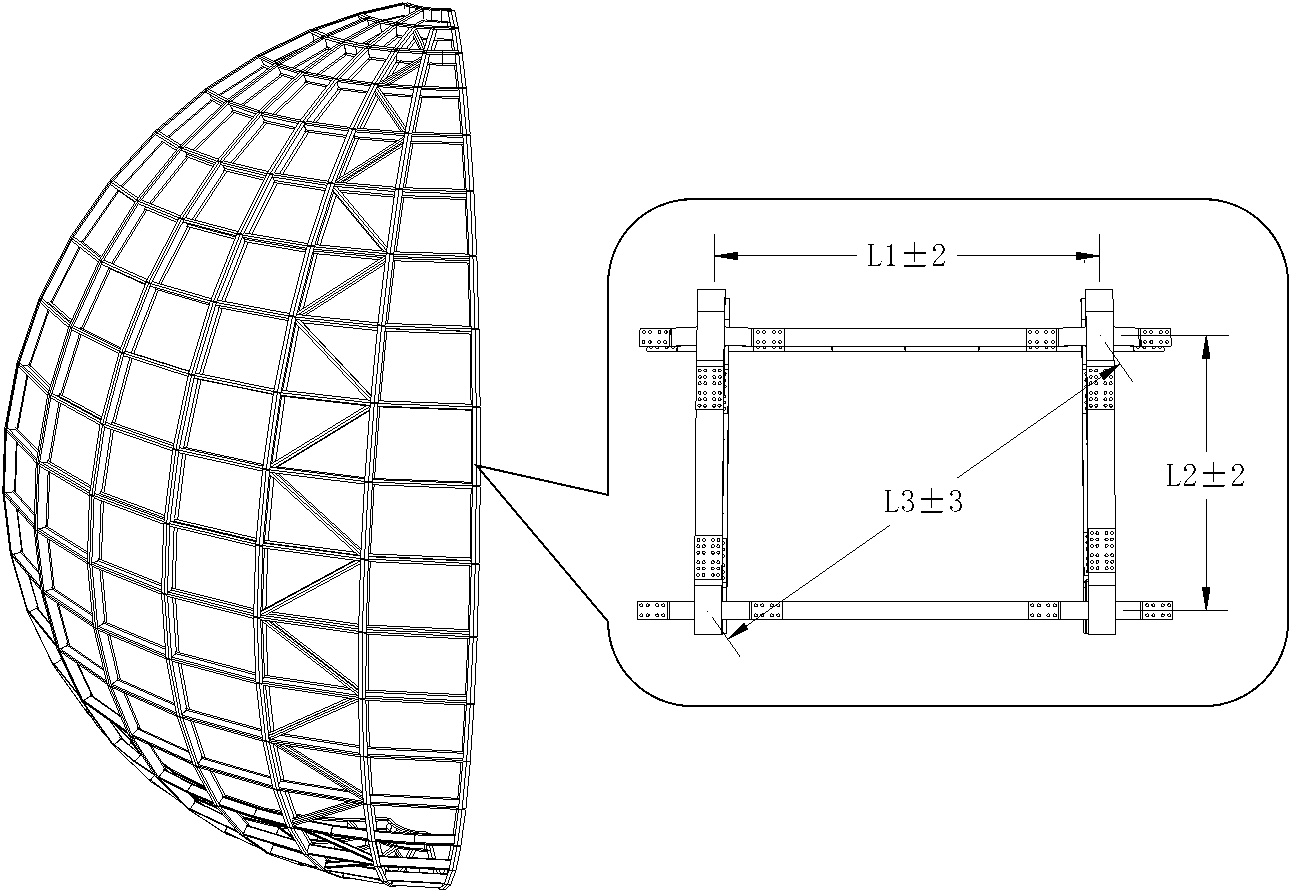}
    \caption{The maximum deviation requirement for the reticulated window of the stainless steel structure.}
    \label{RequirementforReticulatedWindowofSSStructure}       
\end{figure}

The arrangement of 20-inch PMTs around the CD requires a coverage over 75$\%$. Therefore, the installation gap between adjacent PMTs is only 3 mm. The CD SS structure is assembled underground by bolts and the gaps between the bolts and holes are 1 mm on average. Therefore the maximum deviation requirement for the reticulated window of the structure is 3 mm, as shown in Fig. \ref{RequirementforReticulatedWindowofSSStructure}. All these make stringent requirements on the accuracy of the SS structure construction. Two main factors affect the accuracy of the SS latticed shell, the deformation during welding in the manufacturing process and the deviation caused by the bolted joint slipping during the installation process.

To ensure structural strength, all SS components are made by full penetration welding of plates, and the total length of the weld is up to 29,000 m. Since the thermal conductivity of SS is small and the linear expansion coefficient of SS is large, the welding deformation caused by the above reasons must be controlled well.

\begin{figure}[!ht]
    \centering
  \includegraphics[width=0.98\linewidth]{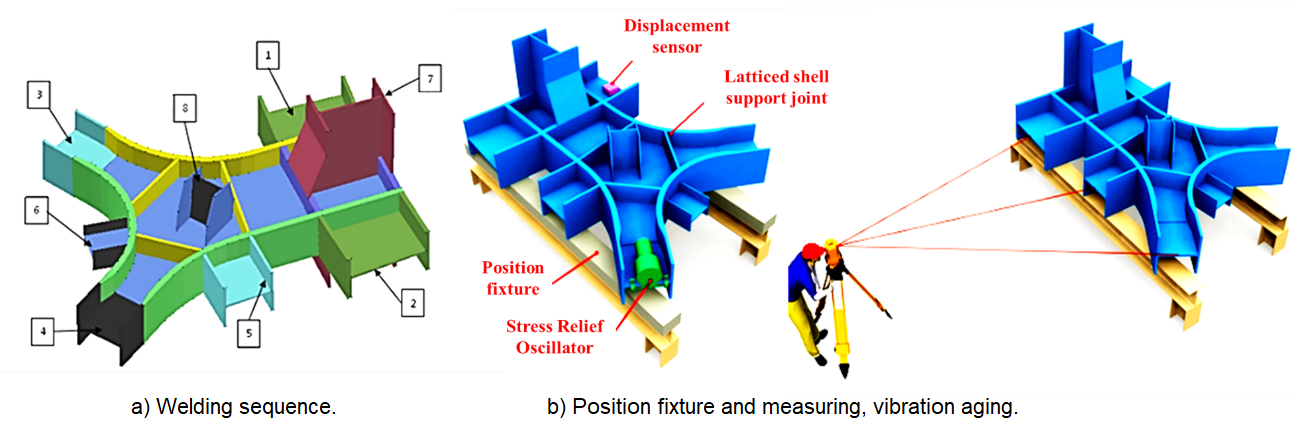}
\caption{The welding of the complex joints of SS structures to control deformation.}
\label{WeldingofComplexJointsofSSStructures}       
\end{figure}

The following measures were taken to reduce welding deformations:
\begin{itemize}
\item laser cutting (t $\leq$ 20 mm) and water jet cutting (t \textgreater  20 mm) were used for SS plates, to improve the material processing accuracy of components and reduce the heat-affected zone;
\item two welding methods, deep penetration Argon arc welding and $CO_{2}$ gas shielded welding, with a reasonable welding sequence have been adopted to reduce the welding heat input and improve the residual stress distribution (see Fig. \ref{WeldingofComplexJointsofSSStructures});
\item special clamping fixtures were used to locate and fix the components, and to ensure their accuracy during assembly;
\item vibration aging after welding was used to reduce the residual stress (see Fig. \ref{WeldingofComplexJointsofSSStructures});
\item all components were checked after welding.
\end{itemize}
The manufacture process was done by the Southeast Space Frame Group Co., ltd. \cite{RefSoutheastSpaceFrameGroup}.

To ensure the stainless steel structure can be installed smoothly and accurately underground, the following measures were taken:
\begin{itemize}
\item all components were pre-assembled at the factory to check the key structure dimensions and match the bolts and connecting holes;
\item hinged bolts were used temporarily as positioning pins when components are installed and connected;
\item the integral positioning fixtures were used for the embedded anchors of 60 supporting columns before pouring concrete at the water pool bottom;
\item after erecting all 60 supporting columns, their positions were checked, and a second round of concrete pouring was performed.
\end{itemize}

\subsection{Special techniques of locking bolt connection}

\begin{figure}[!ht]
    \centering
  \includegraphics[width=0.98\linewidth]{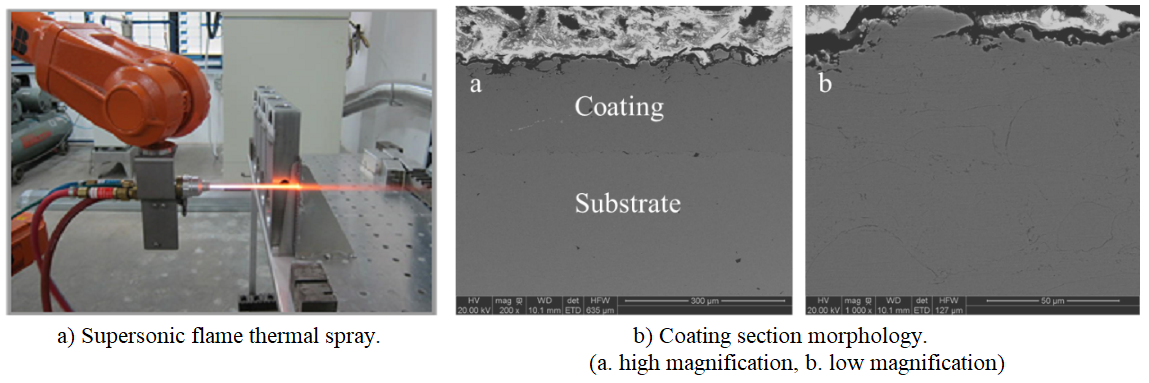}
\caption{Anti-slip treatment on the surface of the connecting plate.}
\label{AntiSlipTreatmentConnectingPlate}       
\end{figure}

To ensure the safety of the installation and the cleanliness of the underground experiment hall, the CD SS structure adopts an all bolt connection design, with a large number of bolts (about 120,000 sets). A friction-critical joint is used because it is safer than a snug-tight joint. The surface of the SS connecting plate was treated with the supersonic flame thermal spray method (see Fig. \ref{AntiSlipTreatmentConnectingPlate}), increasing the joints slip coefficient from 0.2 to 0.5 \cite{RefESSSHSBSRC}.
The scanning electron microscopy results show that the sprayed coating and the original stainless steel substrate are very well bonded, and no obvious interface can be seen. (See fig. 26 b)).

To ensure the reliability and safety of the steel structure connection under the full detector load, 10.9-grade high-strength bolts have been adopted. The bolt material used precipitation hardening stainless steel \cite{RefESMPHSSSSTSLP}, and its low background and permeability also meet the requirements of JUNO.

\begin{figure}[!ht]
    \centering
  \includegraphics[width=0.98\linewidth]{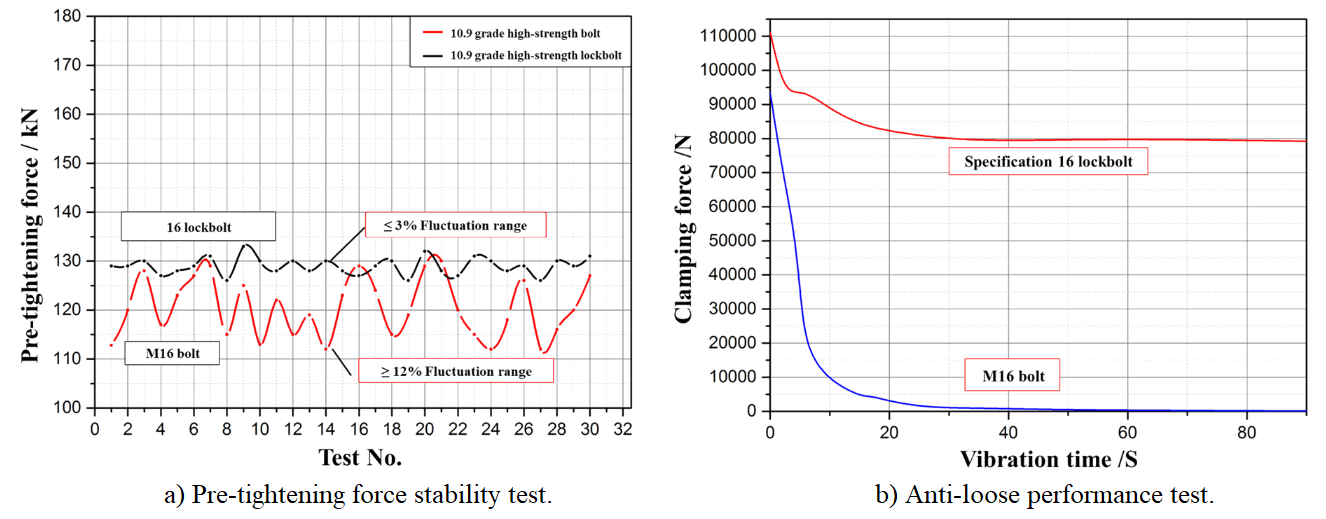}
\caption{High-strength stainless steel lock bolt performance test.}
\label{HighStrengthSSLockBoltPerformancTest}       
\end{figure}

This project uses stainless steel high-strength lock bolt, which has the following advantages:
\begin{itemize}
\item avoid the seizure of thread under high pre-tightening force;
\item fast installation by the special riveting equipment;
\item good consistency of pre-tightening force (see Fig. \ref{HighStrengthSSLockBoltPerformancTest}a);
\item high anti-loose performances (see Fig. \ref{HighStrengthSSLockBoltPerformancTest}b).
\end{itemize}

\section{Assembly and installation underground}

\begin{figure}[!ht]
 \centering
 \includegraphics[width=0.7\linewidth]{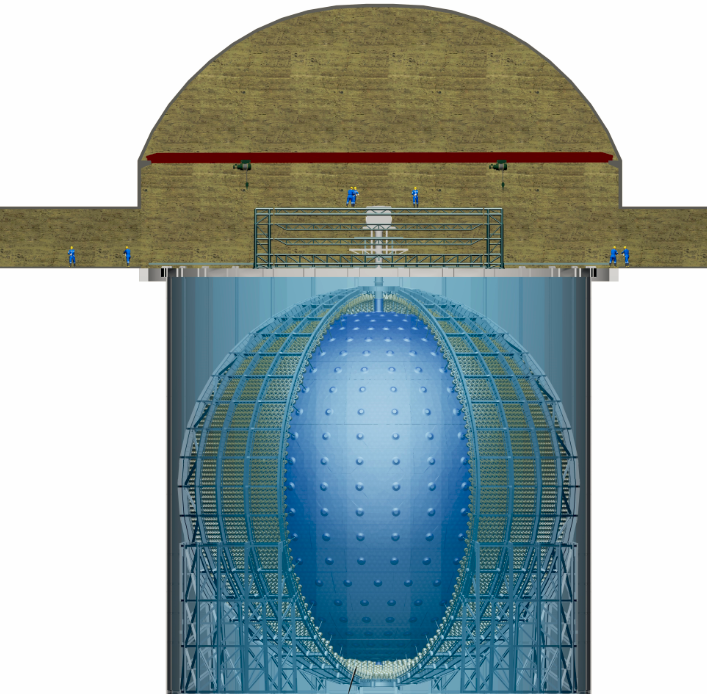}
\caption{The detector in the underground experimental hall.}
\label{TheDetectorintheUndergroundExperimentalHall}       
\end{figure}

After the completion of the underground civil construction, the JUNO detector started the installation phase. Acrylic panels and stainless steel structures were pre-assembled by the manufacturer and then transported to the underground experimental hall by the slope tunnel. The maximum allowed size of the goods to be transported underground on the cable train should be less than 4 m x 3 m x 12 m which is the limitation of tunnel size. The stainless steel pieces will be transported to the top of the water pool and the acrylic panels to the bottom of the water pool using an electrical vehicle. Fig. \ref{TheDetectorintheUndergroundExperimentalHall} shows a drawing of the detector in the underground experimental hall.

\begin{figure}[!ht]
    \centering
  \includegraphics[width=0.98\linewidth]{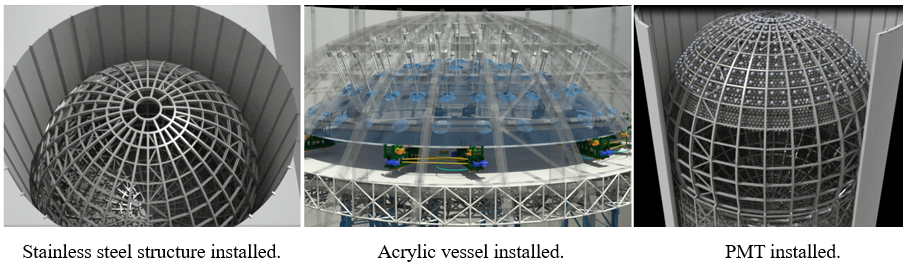}
\caption{Installation sequence of the central detector.}
\label{InstallationSequenceoftheCentralDetector}       
\end{figure}

The stainless steel structure is the first task of the CD installation underground, which is built from bottom to top. Afterwards, the acrylic vessel is constructed on a lifting platform from top to bottom, layer by layer. The PMTs are installed on the stainless steel structure following the acrylic vessel, layer by layer. Fig. \ref{InstallationSequenceoftheCentralDetector} shows the installation sequence.

\subsection{Stainless steel structure installation}

\begin{figure}[!ht]
    \centering
  \includegraphics[width=0.98\linewidth]{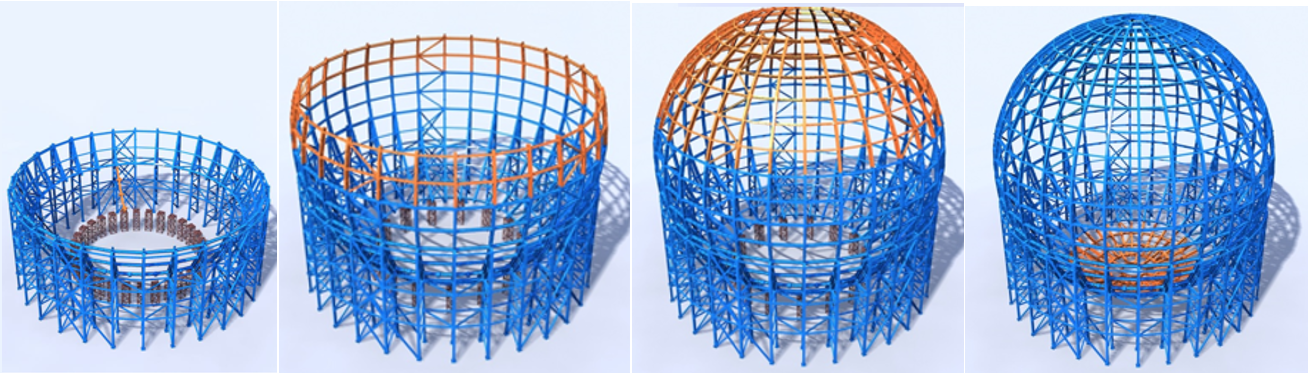}
\caption{Installation flow chart of stainless steel structure.}
\label{InstallationFlowChartofStainlessSteelStructure}       
\end{figure}

No welding is allowed in the experimental hall for the installation of the stainless steel structure, all parts will be bolted in the water pool. In case of any mismatch when bolting, the backup connection plate will be replaced. As a first step, 60 supporting columns will be fixed onto the embedded anchors on the water pool bottom. Afterwards, the latticed shell will be installed in the water pool, layer by layer. To enable the access of acrylic rigging, the four bottom layers will not be completed until the acrylic vessel is finished. Fig. \ref{InstallationFlowChartofStainlessSteelStructure} shows the installation flow chart.

\subsection{Acrylic vessel installation}

\begin{figure}[!ht]
    \centering
  \includegraphics[width=0.98\linewidth]{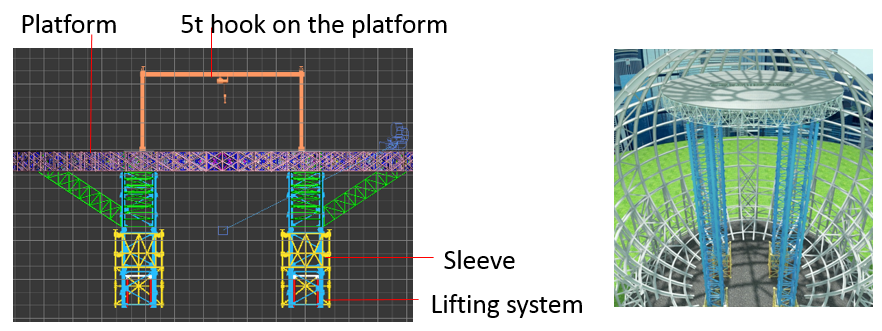}
\caption{Lifting platform for acrylic installation.}
\label{LiftingPlatformforAcrylicInstallation}       
\end{figure}

The lifting platform is assembled and raised to the highest location when the stainless steel structure is finished, and then the acrylic vessel can start its building on the platform. The platform consists of the lifting system and a working area as shown in Fig. \ref{LiftingPlatformforAcrylicInstallation}. The platform has been manufactured by the Tianmu Construction Group Co., Ltd. \cite{RefTianmuConstructionGroup}. The height and diameter of the platform are adjustable to meet the acrylic installation requirement of each layer. A hook on the platform for lifting the acrylic panels is present. Since the stainless steel structure is already in the water pool when the acrylic sphere is under construction, all the acrylic panels will be transported to the bottom of the water pool and then rigged on the lifting platform for the installation. The first layers will be rigged by the 20 t overhead crane in the experimental hall, the hook of the crane can go through the top hole of the stainless steel structure. For the other layers, the acrylic panels will be rigged by the 5 t hook on the lifting platform.

\begin{figure}[!ht]
    \centering
  \includegraphics[width=0.98\linewidth]{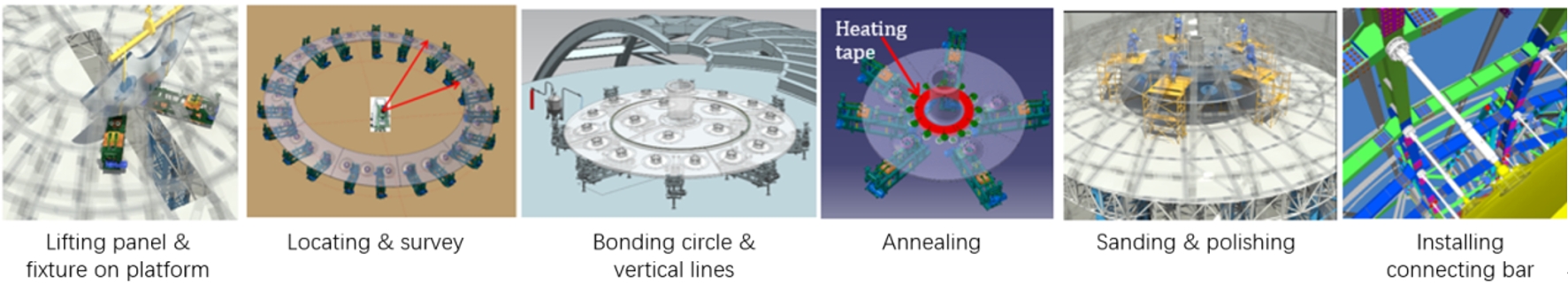}
\caption{Installation steps of each layer for acrylic vessel.}
\label{InstallationStepsofEachLayerforAcrylicVessel}       
\end{figure}

Fig. \ref{InstallationStepsofEachLayerforAcrylicVessel} shows the main installation steps for each layer during the acrylic installation on the platform. The acrylic panels of the first layer will be rigged on the platform and then their placement will be determined using a survey instrument. Once they are in the correct place, they will be fixed in position. The bulk polymerization and annealing will be done as described in section \ref{BulkPolymerizationTechniques}. Afterwards, the bonding area will be sanded, polished, cleaned, and protected by film. The connecting bars between the acrylic vessel and the stainless steel structure will be installed after the annealing process.

\subsection{Background control during installation}

The cleanliness of the inside and outside of the acrylic vessel is required to be, respectively, 10,000 and 100,000 grade during installation. The environmental temperature surrounding the acrylic is required to be 21 ± 1 °C when positioning and bonding the acrylic panels. To realize the above two requirements, four air systems will be arranged on the top of water pool, which provides clean air with a constant temperature to the water pool by air pipes. These pipes will be routed to the three regions including the inside of the acrylic vessel, the space between the PMTs and acrylic vessel, and outer space in the water pool.


The acrylic panel will be protected by films during shipping, positioning, and installation. Before the acrylic panels are bonded, the protecting films in the bonding areas need to be removed, and after being polished, the bonding area will be protected again with a new film, a kind of paper with water soluble glue, which is easier to be cleaned off when finishing the acrylic vessel. Since there is less chance to access the outside of the acrylic vessel after the installation of the PMTs, the films outside will be removed following the PMTs installation. When all the panels installation is completed and the spherical vessel is built, the acrylic vessel will be sealed and and the process will stop, giving the proper time for all possible radon contamination to decay. Simultaneously certain pure water mist will be sprayed to reduce dust inside the spherical vessel and improve air cleanliness. Then the paper of the inside surface will be removed by high pressure ultrapure water, and the inner surfaces will be strictly cleaned to meet the low-background requirement \cite{li2023study}. Finally the vessel will wait for filling.

\section{FOC system}

\subsection{Functions and requirements}

The Filling, Overflowing and Circulating (FOC) system is in charge of LS filling, overflowing or refilling, and online circulation for LS re-purification. Filling the CD with purified LS is the final step in the detector construction. The process must preserve the structural integrity of the detector and the chemical and optical properties of the LS. During detector data taking, the system should also be able to maintain LS pressure and control the LS level variation within 20 cm for detector safety. If the background level and transparency of the LS do not meet the JUNO requirements, a LS online circulation could be implemented for re-purification. To meet the above requirements, a FOC system is designed with an operation lifetime of 20 years.

The requirement of \textsuperscript{210}Pb (daughter of \textsuperscript{222}Rn in the chain of \textsuperscript{238}U) in LS is less than $ 10^{-24} $ g/g, and most \textsuperscript{210}Pb comes from the radon contamination. For the FOC system, the \textsuperscript{210}Pb budget is less than 10$\%$ of the above requirement. Radon can emanate from the material surface into LS, and the emanation rate is proportional to the surface area and the \textsuperscript{238}U concentration. The actual contact surface is largely affected by the surface roughness. Borexino has measured a radon emanation of 4.6 $\mu$Bq/m2 from stainless steel with the roughness of 0.2 $\mu$m and 0.05 ppb \textsuperscript{238}U \cite{RefBorexinoColloaboration2002}. With the help of an atomic force microscope, we found the extra surface area with 0.4 $\mu$m roughness is about 30$\%$. After calculation and evaluation we require the stainless steel used for FOC tanks and pipes to be less than 0.7 ppb \textsuperscript{238}U and 0.4 $\mu$m surface roughness. The stainless steel is customized with low radioactivity and their inner surfaces are electrically polished. 

Another important source of Radon is the underground air, which can diffuse into the LS from any potential leak. The baseline leakage limit of $ 10^{-6} $ mbar\textbullet L/s for Helium is required for each valve and fitting. Additional requirements for the cleanliness of the processing equipment were specified (see Table \ref{RequirementsfortheCleanlinessofRinsedWater}). All the inner surfaces will be washed with detergent and chelating agents, and then rinsed with ultrapure water. The residual particle counts in the rinsed water should meet the required limits in Table \ref{RequirementsfortheCleanlinessofRinsedWater} \cite{RefJUNORCSD}.

\begin{table}[!ht]
\centering
\caption{Requirements for the cleanliness of rinsed water.}
\label{RequirementsfortheCleanlinessofRinsedWater}
\begin{tabular}{ |c|c|c| }
\hline
Particle Size & Surface density & Volume density \\
($\mu$m) & (counts/0.1 $m^2$) & (counts/L) \\
\hline
5 &	179 & 1660 \\
\hline
15 & 27.0 &	250 \\
\hline
25 & 7.88 &	73 \\
\hline
50 & 1.01 &  10 \\
\hline
\end{tabular}
\end{table}

\subsection{System design}

\begin{figure}[!ht]
    \centering
  \includegraphics[width=0.8\linewidth]{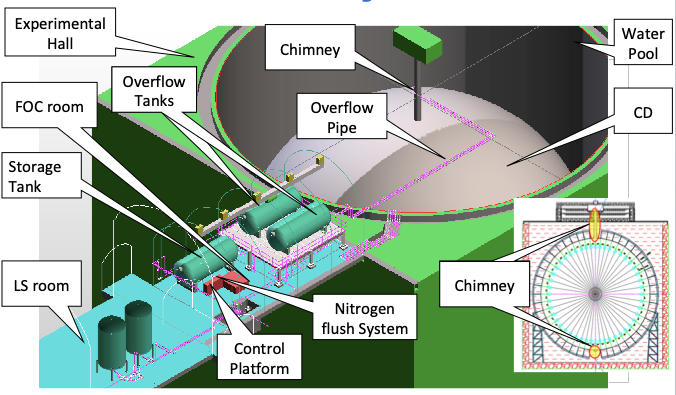}
\caption{The scheme of FOC system.}
\label{TheSchemeofFOCSystem}       
\end{figure}

The FOC includes three LS tanks not only to serve as LS buffers during LS filling and circulating, but also to compensate any volume variation when the temperature changes. Additionally, the system includes pipes, valves, pumps, upper and bottom chimneys, liquid level sensors, flow meters, and control system. The related arrangement of FOC is shown in Fig. \ref{TheSchemeofFOCSystem}.

A radon-free ultrapure nitrogen system was designed to protect the LS from getting in contact with oxygen, moisture, and radon in the FOC tanks, calibration house, and upper chimney. It consists of nitrogen tanks to depressurize radon-free nitrogen before supplying them, mass flow transmitters to monitor the gas flow, bubble bottles to control the gas pressure, and nitrogen pipes. 

\subsubsection{Water and LS filling}

To keep the LS free from air, we need to first replace the air inside the CD with pure water or ultrapure nitrogen gas, and then exchange water or nitrogen with LS. The water exchanging method has been chosen because of less engineering risks. Before filling water into the acrylic spherical vessel, pure water will be used to reduce dust in the air, waiting for possible Rn atoms to decay, thereby reducing the content of dust and Radon inside the CD. Finally, high-pressure rotating nozzles will be used to remove the paper film on the inner surface of the acrylic sphere and to clean it. 

The input pure water can be produced at a rate of about 100 $m^3$/h and will be used to fill the water pool and CD. The design Rn concentration is 10 mBq/$m^3$ and the requirement of \textsuperscript{238}U/\textsuperscript{232}Th concentration is $10^{-15}$ g/g for the pure water to be filled into the water pool. In addition, the water that will be used to fill the CD should undergo a further purification step to meet the more stringent requirements of the detector core. The input flux into those two volumes can be adjusted actively through the proportional valves to keep the water levels balanced and their differences need to be minimized within 50 cm for the safety of the acrylic spherical vessel. The water filling process is finished once the water level reaches 50 cm below the top of the water pool. The whole process will take about two months. 

The flow rate, the liquid levels, and the forces on the connection bars are monitored automatically with high reliability. The pumps and valves are controlled by a programmable logic controller (PLC). The software with the functions of monitoring, controlling and interlock will be developed and implemented based on the distributed Experimental Physics and Industrial Control System (EPICS). The interfaces between this system and the calibration, LS, and water systems will be done through an Input-Output-Controller (IOC) and a database.  

\begin{figure}[!ht]
    \centering
  \includegraphics[width=0.8\linewidth]{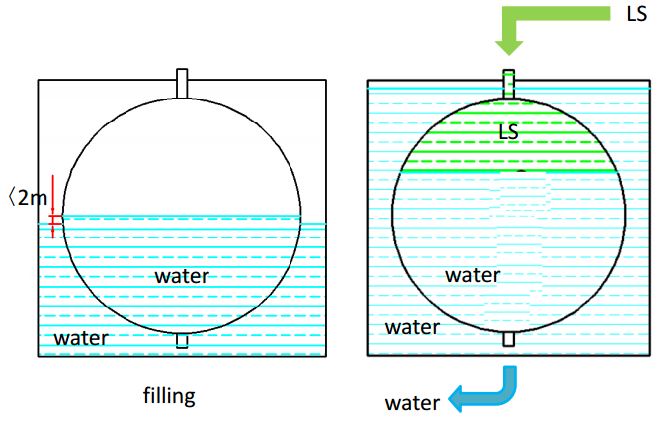}
\caption{The scheme of water-LS exchanging.}
\label{TheSchemeofWaterLSExchanging}       
\end{figure}

After water filling, ultrapure nitrogen needs to be flushed inside the CD to expel the remaining air and reduce the Rn contamination to levels smaller than 10 $\mu$Bq/$m^3$. Before being filled inside the CD, ultrapure LS will be tested with the OSIRIS detector \cite{RefDSJUNOSRPDOSIRIS}, which is a dedicated LS detector to monitor the radioactivity of the input LS. The latter will be filled into the CD through the chimney at the top at a rate of 7 $m^3$/h and water will be pumped out from the bottom of the CD as shown in Fig. \ref{TheSchemeofWaterLSExchanging}. In the end of filling, the water inside the CD will be entirely replaced by the LS, whose level will be 4.5 meters higher than the water level at the water pool (refer to the section \ref{OptimizationoftheChimney}). So, during the process of LS replacement, the height difference between the LS and the water surface in the pool gradually increases from 0 to 4.5 meters.

\subsubsection{Overflow design}
	
The LS thermal volume coefficient is 8.8×$10^{-4}$ / °C so the temperature change will cause high volume changes in a huge volume of LS. Two overflowing tanks and one storage tank with 50 $m^{3}$ volume each and 25 $m^{2}$ cross-section at the center are used to adjust the volume expansion or contraction of the LS inside the CD. They can ensure the safe operation of the CD within the LS temperature range of 21.0 ± 1.4 °C. The two overflowing tanks, located on a supporting platform at a height of 2.3 m, and whose center is at the same level as the CD LS, can passively absorb or refill the LS inside the CD with a level variation within 20 cm when the temperature change is less than 0.4 °C. The storage tank can actively absorb or refill the LS of CD when the temperature changes over 0.4 °C. If the temperature decreases by more than 0.4 °C, the valves and the pumps need to be activated to transfer the LS from the storage tank to overflowing tanks. If the temperature increases over 0.4 °C, the inflated LS can flow from overflowing tanks into the storage tank automatically.

\subsubsection{Circulation design}

To further reduce the impurity, the LS can be circulated and purified online. LS inside the CD will be extracted from the bottom and purified LS will be refilled from the top. With the method of CFD (Computational Fluid Dynamics), some calculations have been done to estimate the circulation efficiency and flow rate. Some useful conclusions are obtained. If there is no temperature difference between new purified LS and old LS in CD, the circulated volume rate is 37$\%$ with the exchange of one CD volume of LS. When the temperature of the new purified LS is 5 °C higher than that of old LS, which can be done to reduce convection, the circulated volume rate is 55$\%$ with the exchange of one CD volume of LS. The optimized LS flow rate of 7 $m^3$/h is determined based on the turbulent motion with such temperature difference.

\section{Summary}

Four main options were proposed for the design of the JUNO CD. The option 1, i.e. an acrylic sphere and stainless steel structure, was finally chosen for the construction of the experiment. An acrylic spherical vessel with an inner diameter of 35.4 m and a thickness of 120 mm, will contain 20 kt of LS. The acrylic vessel has been designed to be highly transparent to scintillation light in order to maximize the light collection of the PMTs on the outside of the sphere. The acrylic vessel is supported by a stainless steel structure with an inner diameter of 40.1 m. It bears almost all loads and supports the acrylic vessel, the PMTs, the front-end electronics, the anti-geomagnetic field coil, and the separating structure for the CD and VETO systems.

The acrylic for JUNO is not a commercially available material. No plasticizer is used in the JUNO acrylic to have good transmittance and mechanical performances for long term operation with acceptable creep and tensile strength aging over 20 years. Moreover, no anti-UV material is used in JUNO acrylic to have better transparency. To reduce the acrylic radioactive background, a special production line was created as follows: 1) filter is added into the beginning pipe of original MMA; 2) the reaction tank is specially made; 3) the mold filled with MMA to produce the acrylic panels is cleaned by pure water in a cleanroom; 4) the surfaces of spherical acrylic panels are polished off to approximately 100 $\mu$m and then packaged by films before shipping to remove the possible \textsuperscript{210}Pb, the daughter of \textsuperscript{222}Rn. It is very important for the safety of the acrylic sphere to control the maximal stress to be less than 3.5 MPa. The acrylic node areas are the stress concentration areas. Therefore, a lot of FEA calculations and tests were made for the acrylic node design. The key and new techniques of locating, bonding, and annealing of the acrylic vessel were studied to reduce the risk and reduce time during installation.

Special welding techniques and fixtures were developed to control the deformation due to welding and obtain the quality and precision required for the stainless steel structure. To ensure the reliability and safety of the steel structure connection when the detector is bearing its full load, the connecting surfaces were treated to get a higher friction coefficient using a special technique. Additionally, the technique of high-strength lock bolts was developed, which can provide consistent and reliable connection of the stainless steel structure and can be installed efficiently with hydraulic guns.

A FOC system was designed to fulfill the functions of Filling, Overflow, and Circulation of LS. The water exchange method was chosen for filling LS. Before filling water into the CD, the dust reduction by water mist and high-pressure nozzle cleaning will be carried out inside CD. Meanwhile Rn content will be reduced because of its decay. To ensure the safety of the CD, both the passive and active overflow control methods were used for LS overflow. The circulation for LS re-purification was optimized to increase efficiency and reduce convection.

The CD aims to realize the radio-purity of $10^{-17}$ g/g for \textsuperscript{238}U and \textsuperscript{232}Th and $10^{-24}$ g/g for \textsuperscript{210}Pb in LS. The materials of the CD were specially produced to reduce their radioactivity, including the acrylic panels, the stainless steel, the glass of PMTs, the LS, etc. Cleaning and radioactive background control are required during installation process.

\section*{Acknowledgments}
\label{sec:Acknowledgments}
We are grateful for the ongoing cooperation from the China General Nuclear Power Group. This work was supported by the Chinese Academy of Sciences, the National Key R$\&$D Program of China, the CAS Center for Excellence in Particle Physics, Wuyi University, and the Tsung-Dao Lee Institute of Shanghai Jiaotong University in China, the Institut National de Physique Nucléaire et de Physique de Particules (IN2P3) in France, the Istituto Nazionale di Fisica Nucleare (INFN) in Italy, the Italian-Chinese collaborative research program MAECI-NSFC, the Fond de la Recherche Scientifique (F.R.S-FNRS) and FWO under the “Excellence of Science – EOS” in Belgium, the Conselho Nacional de Desenvolvimento Científico e Tecnológico in Brazil, the Agencia Nacional de Investigación y Desarrollo in Chile, the Charles University Research Centre and the Ministry of Education, Youth, and Sports in Czech Republic, the Deutsche Forschungsgemeinschaft (DFG), the Helmholtz Association, and the Cluster of Excellence PRISMA+ in Germany, the Joint Institute of Nuclear Research (JINR) and Lomonosov Moscow State University in Russia, the joint Russian Science Foundation (RSF) and National Natural Science Foundation of China (NSFC) research program, the MOST and MOE in Taiwan, the Chulalongkorn University and Suranaree University of Technology in Thailand, and the University of California at Irvine in USA.

\bibstyle{spphys}       
\bibliography{Ref}   

\end{document}

%% file: fullname_231009.tex
\author[6,5]{Angel Abusleme}
\author[44]{Thomas Adam}
\author[65]{Shakeel Ahmad}
\author[65]{Rizwan Ahmed}
\author[54]{Sebastiano Aiello}
\author[65]{Muhammad Akram}
\author[65]{Abid Aleem}
\author[47]{Tsagkarakis Alexandros}
\author[20]{Fengpeng An}
\author[22]{Qi An}
% ...
\author*[]{(JUNO collaboration)}

\email{juno\_pub\_comm@juno.ihep.ac.cn (The JUNO Collaboration)}

\email{maxy@ihep.ac.cn (Xiaoyan Ma)}

\email{hengyk@ihep.ac.cn (Yuekun Heng)}

% ...
\affil[5]{Millennium Institute for SubAtomic Physics at the High-energy Frontier (SAPHIR), ANID, Chile}
\affil[6]{Pontificia Universidad Cat\'{o}lica de Chile, Santiago, Chile}
% ...
\affil[20]{Sun Yat-Sen University, Guangzhou, China}
% ...
\affil[22]{University of Science and Technology of China, Hefei, China}
% ...
\affil[44]{IPHC, Universit\'{e} de Strasbourg, CNRS/IN2P3, F-67037 Strasbourg, France}
% ...
\affil[47]{III. Physikalisches Institut B, RWTH Aachen University, Aachen, Germany}
% ...
\affil[54]{INFN Catania and Dipartimento di Fisica e Astronomia dell Universit\`{a} di Catania, Catania, Italy}
% ...
\affil[65]{Pakistan Institute of Nuclear Science and Technology, Islamabad, Pakistan}
% ...
